\documentclass[fleqn,]{wlscirep}
\usepackage[utf8]{inputenc}
\usepackage{algorithm}
\usepackage{algpseudocode}
\usepackage[T1]{fontenc}
\usepackage{bm}
\usepackage{float}
\usepackage{multirow}
\usepackage{wrapfig} 
\usepackage{makecell} 
\usepackage{quantikz}
\usepackage{subcaption}
\usepackage{tabularx}
\usepackage{graphicx}
\usepackage{pdflscape}
\usepackage{caption}
\usepackage{xcolor}
\usepackage{longtable}
\usepackage[normalem]{ulem}

\title{Resource efficient Sample Based Quantum Simulation through Fragmentation and Virtual Partitioning of QPU }

\author[1]{\small R. Pradeep}
\author[1]{\small Anurag Chaudhary}
\author[1]{\small Ashish Kumar Patra}
\author[1]{\small Anurag K.S.V.}
\author[1,2]{\small Ruchika Bhat}
\author[1]{\small Sai Shankar P.}

\author[, 1]{\small G. Jaiganesh \thanks{(Corresponding Author) email: jaiganesh@qclairvoyance.in, drjaiganesh15@gmail.com}}

\affil[1]{\small Quantum Assisted Drug Discovery Group, Qclairvoyance Quantum Labs, Secunderabad, TG 500094, India.}
\affil[2]{\small The University of Arizona, Tucson, AZ 85721, USA.}

\begin{abstract}
Present-day sample-based quantum algorithms within quantum-centric supercomputing workflows require substantial classical compute resources for post-processing, large circuit depths to produce chemically relevant samples, and long QPU times to execute the large number of circuits needed to simulate molecules of chemically relevant size. In this work, we introduce FMO-QVM-HCI-HSQD, a resource-efficient sample-based framework that tackles these challenges by using fragment-based methods to break a large problem into smaller pieces, running the fragments in parallel by allocating virtual regions of the QPU, and reducing circuit depth by adopting a half-qubit ansatz whose samples seed an HCI expansion. We demonstrate the capabilities of our algorithm through experiments on \textbf{\texttt{ibm\_marrakesh}} and \textbf{\texttt{ibm\_boston}}  for the water trimer and the carmofur--M\textsuperscript{pro} complex with up to 52 qubits, comparing the results with classical benchmarks such as FMO-CCSD(T), FMO-CASCI and FMO-HCI. For the water trimer, which requires 14 spin orbitals for monomers and 28 spin orbitals for dimers, our method reduces the two-qubit depth by 39\% compared to FMO-QVM-SQD and the QPU time by 82\% compared to sequential FMO-SQD, and accesses 3\% of the determinant subspace compared to 40\% for FMO-SQD, while achieving more accurate energies. For the carmofur--M\textsuperscript{pro} complex, which requires 16 spin orbitals for monomers and 32 spin orbitals for dimers, our method reduces the two-qubit depth by 47\% and the QPU time by 20\% compared to FMO-QVM-SQD, and accesses 0.007\% of the determinant subspace compared to 40\% for FMO-QVM-SQD, while remaining within chemical accuracy. Extending the carmofur--M\textsuperscript{pro} active space in the 6-31G basis to 52 spin orbitals for monomers and 104 spin orbitals for dimers, sampled on 26 and 52 qubits respectively, the total energy deviates from FMO-HCI by only 1.22~mHa, within chemical accuracy. These results show that combining fragmentation, virtual QPU partitioning and half-qubit sampling makes sample-based quantum simulation of chemically relevant systems feasible on current hardware.
\end{abstract}

\begin{document}

\flushbottom
\maketitle

\thispagestyle{empty}


\section{Introduction}
\label{sec:introduction}

Molecular simulation is among the most promising areas in which quantum computers are expected to make a practical difference. Considerable effort has gone into designing quantum algorithms that can outperform their classical counterparts in the large-scale strongly correlated regime without demanding resources far beyond what current devices provide. Algorithms in the Fault-tolerant Quantum Computation (FTQC) category, including Quantum Phase Estimation \cite{kitaev1995abelian_qpe, aspuruguzik2005_simulated, scalable_qc_qpe_vqe} and Quantum Singular Value Transformation \cite{Gily_n_2019_qsvt}, achieve near-optimal precision scaling but require circuit depths well above what present Noisy Intermediate-Scale Quantum (NISQ) hardware can execute reliably \cite{Preskill_2018}. Variational schemes such as the Variational Quantum Eigensolver were introduced to operate within those constraints \cite{Peruzzo2014_vqe, Tilly2022_VQE_Review}, replacing deep coherent evolution with a shallow parametrized circuit and a classical optimization loop. The difficulty is that this loop does not scale gracefully. For sufficiently expressive ansatze the gradients of the cost function concentrate exponentially with system size, producing barren plateaus in which the optimizer has no reliable descent direction \cite{mcclean2018barren, Larocca2025_BP_review}. Hardware noise compounds the problem independently, flattening the landscape further and biasing the estimator that guides each update \cite{Wang2021_NIBP}. Sample-based quantum methods were introduced to avoid this failure mode altogether by removing the variational loop from the algorithm \cite{kohda2022quantum, kanno2023qsci, sqd_first_paper}.

Sample-based methods represent a more favourable division of labour \cite{kohda2022quantum, kanno2023qsci, sqd_first_paper}. The quantum processor is used only as a sampler of electronic configurations, and the energy is recovered by classical diagonalization within the sampled subspace, which removes the variational optimization loop and with it much of the shot-noise and measurement burden that limits expectation-value estimation \cite{PhysRevResearch_2022_Gonthier}. The framework has since been extended to excited states \cite{Barison2025_QCExcitedStates_SQD}, Krylov subspace constructions \cite{Yu2025_SBKrylov, Yoshioka2025KrylovDiagonalization, asthana2025quantumkrylovalgorithmusing}, randomized formulations with convergence guarantees \cite{piccinelli2025_sqdrift}, closed-loop orbital optimization \cite{shirakawa2025closedloopcalculationselectronicstructure}, and reduced-width encodings that halve the register requirement \cite{McFarthing2026_hcihsqd, Yoshida2026_dociqsci}, and now underpins quantum-centric supercomputing workflows for chemistry and materials \cite{Alexeev2024_QCSMaterials_sqd}.

Sampling based algorithms overcomes the barren plateau problem but sampling does not remove the cost of preparing a state whose support already covers the chemically relevant configurations. For realistic active spaces this requires one qubit per spin orbital and an ansatz deep enough to generate the correct correlation structure \cite{lucj_ansatz}, and the quality of the resulting sample set degrades as either quantity grows on noisy hardware. Qubit count and circuit depth therefore control whether a sample-based calculation returns configurations of any use on present devices, and both remain binding constraints even where the method itself is sound \cite{critical_lims_sqd_2025}. A third quantity, the QPU time consumed once many independent samples must be collected, decides whether the calculation is practical at all.

Our main aim is to combine resource-efficient workflow with sample-based algorithms to enable quantum simulation of molecular ground states efficiently and based on this we design following methods to achieve this goal. 

The qubit count is bounded by fragmentation. Embedding methods replace the global wavefunction with a set of subsystem problems and rebuild the total energy from fragment contributions, reintroducing inter-fragment coupling through embedding potentials, density matrix matching, or effective fragment Hamiltonians \cite{Sun_2016_quantum_embedding_theories}. The established methods are DMET and its multireference extensions \cite{Knizia2012_DMET, wouters2016dmets_guide, Pham2018DMET_laura, Ai_2022_DMET_ROHF_SIM}, Divide and Conquer \cite{Nakai2023DC}, bootstrap embedding \cite{ye2020bootstrap}, localized and wavefunction-based embedding \cite{Hermes2020VLASSCF, ewf_embedding_2022, nusspickel_ewf_2023}, coupled cluster downfolding \cite{Bauman2023CCDownfolding}, and the Fragment Molecular Orbital framework \cite{fmo_fedorov_2012, EFMO_Sattasathuchana_2024}. Coupling such decompositions to a quantum solver is equally well established, from early problem-decomposition proposals \cite{yamazaki2018_practical_appl} through DMET \cite{Kawashima2021_dmet_vqe_ion, dmet_sqd, Patra2025_DMETSQD_ligand, ksv2026}, bootstrap embedding \cite{Liu2023Bootstrap, Bierman_2026_QBE_SQD}, localized orbital embedding \cite{Otten2022Localized_laura, shajan2026_ewf_protein}, divide-and-conquer circuits \cite{fujii_2022_deep_vqe}, many-body expansions \cite{jinlong_2023_mbe, Xu2024MBE_VQE_Deflation}, downfolding and flow embedding \cite{bauman2025coupledclusterdownfoldingtheory, kowalski2024resourceadaptivequantumflowalgorithms}, and FMO \cite{FMO_VQE_Lim_2024}, with applications reaching protein-ligand and metal-protein systems \cite{Hardikar2024, Bowling2025_ProteinLigand_FragQC}. We introduce no new decomposition. Our contribution is to show  what a fragmented workload permits once the solver and the execution layer are built around it.
We adopt FMO in our algorithm because its partitioning is  physically transparent and easy to align with a qubit budget. A fragment is a monomer and two monomers taken together forms a dimer. The many-body expansion is explicit and truncatable, so FMO provides  systematic control of accuracy rather than a black box. The environment enters through a self-consistent Coulomb field built from the remaining fragment densities, so each fragment is an independent embedded problem and the correlated solver at fragment level can be replaced without disturbing the rest of the pipeline.  Since the fragment subproblems are mutually independent, they can be  executed concurrently as described below possible at all.
\\

The circuit depth is reduced at the fragment level. Even within a fragment active space, populating the configurations needed for chemical accuracy demands entangling depth that noisy hardware struggles to sustain \cite{Preskill_2018, superconducting_quantum_engineer}, which is why chemically motivated low-depth ansatze such as the local unitary cluster Jastrow construction have become the standard sampling circuits \cite{lucj_ansatz, Matsuzawa2020_JastrowLowDepth, Motta2021_lowrank_quCCSD}. We sample instead from a half-qubit formulation of the ansatz \cite{McFarthing2026_hcihsqd, Yoshida2026_dociqsci}, which halves the register and reduces entangling-gate count while preserving the configuration support the subsequent diagonalization requires. Selected-CI diagonalization \cite{sci_evangelista_1} is then carried out with heat-bath CI \cite{Holmes2016_HCI, Sharma2017_SHCI, Li2018_FastSHCI}, which expands the sampled set along the largest Hamiltonian couplings and lets a comparatively sparse sample yield a subspace of the quality needed for chemical accuracy. Together these give the HCI-HSQD solver used at every fragment.
\\

The QPU time is reduced by sharing the device. NISQ processors are scarce, and an individual workload typically occupies only a fraction of the qubits on a device, leaving the remainder idle. Quantum multi-programming addresses this by executing independent programs concurrently on disjoint portions of the same QPU, improving utilization and throughput \cite{Das2019}, provided the non-uniform reliability of physical qubits and couplers is accounted for. Later work built this hardware awareness into partitioning and mapping directly. QuCloud partitions the coupling graph by community detection and schedules workloads onto the resulting regions \cite{QuCloud}, QuMC incorporates topology, calibration data, and crosstalk characterization into allocation \cite{Niu2023}, and QuCloud+ extends topology-aware mapping to 2D and 3D architectures \cite{QuCloudPlus}. These results establish that topology and qubit quality matter when a device is shared, but they treat sharing as a partitioning and scheduling problem rather than as a resource abstraction. Quantum Virtual Machine (QVM) raise this a level by presenting portions of a QPU as independent virtual execution resources. HyperQ introduced the model, multiplexing several virtual machines onto one processor across space and time while compiling each program independently of the full device configuration~\cite{tao2025quantum}. However, the virtual boundaries in this approach are largely static and cannot track how hardware quality varies across the chip and over time. DynQ addresses this by deriving execution regions from a quality-weighted representation of the processor and separating offline region discovery from online allocation, so the virtual structure adapts to drifting hardware while runtime allocation remains lightweight~\cite{DynQ}.
\\

FMO produces many mutually independent circuits, which is the kind of workload this machinery is designed for. While the half-qubit reduction allows many fragment circuits to fit comfortably inside a single high-quality atomic region, fragments whose requirements exceed a single region's capacity are dynamically accommodated through multi-region composition~\cite{DynQ}. We therefore use DynQ's dynamic, quality-weighted QVM as the spatial foundation of the execution layer, discovering physical execution regions, allocating fragment circuits to atomic or composed regions, and running independent fragments concurrently across disjoint regions. We extend the allocator with temporal reuse of occupied regions inspired by HyperQ's time multiplexing~\cite{tao2025quantum} through a Barrier-Reset-Barrier (BRB) scheduling step. Region-aware compilation, circuit combination, execution, and result recovery are handled transparently, so fragment circuits share a physical QPU without the application managing topology or allocation itself. The wall-clock QPU time for the full molecule then scales with the concurrent capacity of the device rather than strictly with the number of fragments.
\\

Combining these elements, we propose FMO-QVM-HCI-HSQD, a resource-efficient framework for sample-based quantum simulation of molecules. The main contributions of this work are:
\begin{enumerate}[label=(\roman*)]
    \item A framework, FMO-QVM-HCI-HSQD, that combines fragment molecular orbital decomposition, a quality-aware quantum virtual machine, and half-qubit sample-based diagonalization.
    \item An extension of DynQ's region allocation with temporal reuse through a BRB sequence, so that all fragment circuits execute within a single job.
    \item A hardware demonstration on IBM Heron processors for the water trimer and the carmofur--M\textsuperscript{pro} PL complex, with fragments of up to 52 qubits, compared against FMO-FCI, FMO-CASCI, FMO-CCSD(T), and FMO-HCI references.
\end{enumerate}

The rest of the paper is organized as follows. Section~\ref{sec:theoretical_background} reviews FMO, SQD, HCI-HSQD, and the QVM. Section~\ref{sec:methodology} describes the fragmentation and the QVM integration. Section~\ref{sec:results_and_discussion} presents the results, including the large-scale experiments, and Section~\ref{sec:conclusion} concludes with the limitations of this study.
\section{Theoretical Background}
\label{sec:theoretical_background}

\subsection{Fragment Molecular Orbital Method}
The fragment molecular orbital (FMO) method
\cite{Kitaura1999_FMO, fmo_fedorov_2012} approximates the electronic
structure of a large system by partitioning it into small, non-overlapping
fragments and solving each fragment, together with each pair of fragments, in
the electrostatic field generated by all the others. In its two-body form
(FMO2), which we adopt throughout this work, the calculation proceeds in four
stages: fragmentation of the system, a self-consistent monomer stage, a dimer
stage, and assembly of the total energy from the fragment results. The scheme
has been applied to systems of thousands of atoms
\cite{fmo_fedorov_2012, EFMO_Sattasathuchana_2024} and has previously been
combined with quantum solvers at the fragment level \cite{FMO_VQE_Lim_2024}.
`
The system is first partitioned into $N_{\mathrm{f}}$ disjoint fragments
(monomers), labelled $X$, $Y$, $Z$. The eigen states and eigen values of the Hamiltonian for monomer $X$
are obtained by solving
\begin{equation}
  \hat{\mathcal{H}}_{X}\,\Phi_{X} = \varepsilon_{X}\,\Phi_{X},
  \label{eq:mono-schro}
\end{equation}
where the monomer Hamiltonian, embedded in the electrostatic potential of the
remaining $N_{\mathrm{f}}-1$ fragments, is
\begin{equation}
  \hat{\mathcal{H}}_{X}
  = \sum_{p \in X}
    \left[
      -\tfrac{1}{2}\nabla_p^{2}
      - \sum_{A} \frac{Z_A}{\lvert \mathbf{x}_p - \mathbf{R}_A \rvert}
      + \sum_{Y \neq X}^{N_{\mathrm{f}}}
        \int \mathrm{d}\mathbf{x}'\,
        \frac{\varrho_{Y}(\mathbf{x}')}{\lvert \mathbf{x}_p - \mathbf{x}' \rvert}
    \right]
  + \sum_{p<q \in X}
      \frac{1}{\lvert \mathbf{x}_p - \mathbf{x}_q \rvert} .
  \label{eq:mono-ham}
\end{equation}
Here $Z_A$ and $\mathbf{R}_A$ are the nuclear charges and positions in which index A runs over nuclei of fragment X only, and
$\varrho_{Y}(\mathbf{x}')$ is the electron density of fragment $Y$. The third
term inside the bracket is the electrostatic embedding potential, which
couples each electron of fragment $X$ to the charge distribution of every
other fragment at the mean-field level. Because this potential is built from
the fragment densities, which in turn depend on the embedded orbitals, the
monomer calculations are iterated to mutual self-consistency in the
self-consistent-charge (SCC) procedure \cite{Kitaura1999_FMO, fmo_fedorov_2012}.
Convergence of the SCC loop yields the embedded monomer energies
$\varepsilon_{X}$ and densities $\varrho_{X}$.

Each fragment pair (dimer) $XY$ is then solved in the same manner,
\begin{equation}
  \hat{\mathcal{H}}_{XY}\,\Phi_{XY} = \varepsilon_{XY}\,\Phi_{XY},
  \label{eq:dim-schro}
\end{equation}
with the dimer Hamiltonian now embedded in the potential of the remaining
$N_{\mathrm{f}}-2$ fragments,
\begin{equation}
  \hat{\mathcal{H}}_{XY}
  = \sum_{p \in X,Y}
    \left[
      -\tfrac{1}{2}\nabla_p^{2}
      - \sum_{A} \frac{Z_A}{\lvert \mathbf{x}_p - \mathbf{R}_A \rvert}
      + \sum_{Z \neq X,Y}^{N_{\mathrm{f}}}
        \int \mathrm{d}\mathbf{x}'\,
        \frac{\varrho_{Z}(\mathbf{x}')}{\lvert \mathbf{x}_p - \mathbf{x}' \rvert}
    \right]
  + \sum_{p<q \in X,Y}
      \frac{1}{\lvert \mathbf{x}_p - \mathbf{x}_q \rvert} ,
  \label{eq:dim-ham}
\end{equation}
where $Z$ runs over the fragments other than $X$ and $Y$. The dimer is solved
exactly once using the converged environment densities from the monomer stage, giving the
embedded dimer energies $\varepsilon_{XY}$.

In the final stage the fragment results are pieced together to obtain the total
energy of the system. At the restricted Hartree--Fock (RHF) level the FMO2
energy is written as a monomer contribution plus a two-body correction,
\begin{equation}
  E^{\mathrm{RHF}}_{\mathrm{FMO2}} = E^{\mathrm{RHF}}_{\mathrm{FMO1}} + \Delta E^{\mathrm{RHF}}_{\mathrm{FMO2}},
  \label{eq:fmo2-rhf}
\end{equation}

\begin{equation}
  E^{\mathrm{RHF}}_{\mathrm{FMO1}} = \sum_{X}^{N_{\mathrm{f}}} \varepsilon_{X},
  \label{eq:fmo1-rhf}
\end{equation}

\begin{equation}
  \Delta E^{\mathrm{RHF}}_{\mathrm{FMO2}} = \sum_{X<Y}^{N_{\mathrm{f}}} \bigl( \varepsilon_{XY} - \varepsilon_{X} - \varepsilon_{Y} \bigr).
  \label{eq:dfmo2-rhf}
\end{equation}
$\mathcal{O}(N_{\mathrm{f}})$
The FMO1 term, Eq.~\eqref{eq:fmo1-rhf}, is the sum of the embedded monomer
energies, in which each fragment is polarized by the field of its environment.
The two-body term, Eq.~\eqref{eq:dfmo2-rhf}, is the interfragment interaction
energy. Subtracting the monomer energies $\varepsilon_{X}$ and $\varepsilon_{Y}$,
already counted in the FMO1 sum, isolates the electrostatic, exchange, and
charge-transfer contributions that arise only when fragments $X$ and $Y$ are
treated together. The total number of fragments scale quadratically ($\mathcal{O}(N^2)$)
\cite{fmo_fedorov_2012, EFMO_Sattasathuchana_2024}.

Electron correlation is incorporated by adding a correlation energy to the
RHF reference, evaluated with the same two-body fragmentation. The total FMO2
energy of a correlated method becomes
\begin{equation}
  E_{\mathrm{FMO2}}
  = E^{\mathrm{RHF}}_{\mathrm{FMO2}}
  + E^{\mathrm{corr}}_{\mathrm{FMO2}},
  \label{eq:fmo2-total}
\end{equation}
where the correlation energy is accumulated in the same monomer-plus-pair form
as the reference energy,
\begin{equation}
  E^{\mathrm{corr}}_{\mathrm{FMO2}} = E^{\mathrm{corr}}_{\mathrm{FMO1}} + \Delta E^{\mathrm{corr}}_{\mathrm{FMO2}},
  \label{eq:fmo2-corr}
\end{equation}

\begin{equation}
  E^{\mathrm{corr}}_{\mathrm{FMO1}} = \sum_{X}^{N_{\mathrm{f}}} \varepsilon^{\mathrm{corr}}_{X},
  \label{eq:fmo1-corr}
\end{equation}

\begin{equation}
  \Delta E^{\mathrm{corr}}_{\mathrm{FMO2}} = \sum_{X<Y}^{N_{\mathrm{f}}} \bigl( \varepsilon^{\mathrm{corr}}_{XY} - \varepsilon^{\mathrm{corr}}_{X} - \varepsilon^{\mathrm{corr}}_{Y} \bigr).
  \label{eq:dfmo2-corr}
\end{equation}
Here $\varepsilon^{\mathrm{corr}}_{X}$ and $\varepsilon^{\mathrm{corr}}_{XY}$ are the
correlation energies of monomer $X$ and dimer $XY$, each obtained as the
difference between the correlated and RHF energies of the corresponding
embedded fragment. The correlated solve is performed on each monomer once its
embedded SCF has converged, and on each dimer once its embedded SCF has
converged. In this work we use SQD \cite{sqd_first_paper, kanno2023qsci} and
HCI-HSQD \cite{McFarthing2026_hcihsqd, Holmes2016_HCI} to obtain the
correlated energies of monomers and dimers with an active space set by number of HOMO-LUMO orbitals we choose. Orbitals outside this active space are treated at the RHF level.

For a system partitioned into $N$ fragments, the two-body expansion requires
a separate solution of each fragment (monomer) and of every distinct pair of
fragments (dimer). \textsc{OpenFMO} therefore generates $N$ monomer
Hamiltonians and $\binom{N}{2}$ dimer Hamiltonians, so that the total number
of fragment Hamiltonians to be solved is
\begin{equation}
  N_{\mathrm{H}}
  = N + \binom{N}{2}
  = N + \frac{N(N-1)}{2}
  = \frac{N(N+1)}{2}
  = \mathcal{O}(N^{2}).
  \label{eq:num-hamiltonians}
\end{equation}




\subsection{Sample-Based Quantum Diagonalisation (SQD)} \label{sub-sec: Theory Sub-Heading 2}
Sample-Based Quantum Diagonalisation was introduced in
Ref.~\cite{sqd_first_paper}. SQD combines the QSCI quantum sampling scheme
\cite{kohda2022quantum, kanno2023qsci} with the S-CoRe (self-consistent
configuration recovery) error-mitigation technique to repair noisy
configurations. The recovered configurations define an initial subspace that is
subsequently expanded, and the Hamiltonian is projected on this expanded subspace and diagonalized to obtain the ground-state energy. The steps of the
algorithm are described briefly below; for a detailed discussion of the
procedure and its behaviour in an embedded fragment setting we refer the reader
to Refs.~\cite{dmet_sqd, Patra2025_DMETSQD_ligand, ksv2026}.

\begin{enumerate}

\item \textbf{State preparation.} A Hartree--Fock reference is acted on by the
LUCJ ansatz~\cite{lucj_ansatz}, $\ket{\psi_{\mathrm{init}}} =
\hat{U}_{\mathrm{LUCJ}}(\theta_1,\dots,\theta_n)\ket{\psi_{\mathrm{HF}}}$,
with the parameters fixed \emph{a priori} from the CCSD amplitudes
$t_{i}^{a}$ and $t_{ij}^{ab}$~\cite{lucj_ansatz, Motta2021_lowrank_quCCSD}. No
variational optimization loop is required, the ansatz need only carry
appreciable overlap with the true ground state. Hardware comparisons against
deeper unitary coupled-cluster circuits indicate that this shallow, Jastrow-based
construction helps the sampled distribution usable on current
devices.

\item \textbf{Computational-basis sampling.} Projective measurement in the
$\hat{\sigma}_z$ basis yields $d$ configurations
$\mathcal{S}_{\mathrm{samp}} = \{\ket{\varphi_1},\dots,\ket{\varphi_d}\}$,
each factorizing as $\ket{\varphi_i} = \ket{\alpha_i}\otimes\ket{\beta_i}$.
For a closed-shell singlet the physical sector is fixed by
\begin{equation}
  N_\alpha + N_\beta = N_{\mathrm{elec}}, \qquad N_\alpha - N_\beta = 0 ,
  \label{eq:sym}
\end{equation}
and device noise generates samples that violate both constraints. For sufficiently large circuits, this failure of sampled bit strings to follow the symmetry requirements becomes very common. 
\item \textbf{Self-consistent configuration recovery (S-CoRe).} Given the
average spin-orbital occupancies $n_{p\sigma} = \bra{\Psi} \hat{a}^{\dagger}_{p\sigma}
\hat{a}_{p\sigma} \ket{\Psi}$, symmetry-broken samples are repaired by
probabilistically flipping occupations until Eq.~\eqref{eq:sym} is
satisfied~\cite{sqd_first_paper}.
Distributing this over $K$ batches produces recovered subspaces
$\{\mathcal{S}^{(b)}_{\mathrm{CR}}\}_{b=1}^{K}$ that respect the target
symmetries while remaining close to the sampled distribution.

\item \textbf{Subspace construction and projected diagonalization.} The
projection space is \emph{not} $\mathcal{S}^{(b)}_{\mathrm{CR}}$ itself. The
unique $\alpha$ and $\beta$ strings are first merged,
$\mathcal{P}_{\mathrm{unique}} = \mathcal{A}_{\mathrm{unique}} \cup
\mathcal{B}_{\mathrm{unique}}$, and the subspace is taken as the self
tensor product
\begin{equation}
  \mathcal{S}_{\mathrm{proj}} = \mathcal{P}_{\mathrm{unique}} \otimes
  \mathcal{P}_{\mathrm{unique}}, \qquad
  \dim \mathcal{S}_{\mathrm{proj}} = |\mathcal{P}_{\mathrm{unique}}|^{2},
  \label{eq:sproj}
\end{equation}
which enlarges the space to include alternative pairings of the sampled spin
sectors, motivated by the observation that if $\ket{\alpha}\otimes\ket{\beta}$
carries weight in the ground state then recombinations drawn from the same
sectors generally do as
well~\cite{sqd_first_paper, kanno2023qsci, Patra2025_DMETSQD_ligand}.
Projecting with $\hat{P}_{\mathcal{S}_{\mathrm{proj}}} = \sum_{x \in
\mathcal{S}_{\mathrm{proj}}} \ket{x}\bra{x}$ and diagonalizing
$\hat{H}_{\mathcal{S}_{\mathrm{proj}}} = \hat{P}\hat{H}\hat{P}$ by Davidson's
method~\cite{Davidson1975_iterative_eigensolver} returns $E^{(b)}$ which is a variational upper bound to the true
ground-state energy~\cite{MacDonald1933_RayleighRitz} of the hamiltonian in the chosen active space.

\item \textbf{Self-consistent update and convergence.} The running estimate is
$E = \min_{b} E^{(b)}$, and the occupancies are refreshed from the batch
solutions,
\begin{equation}
  \bar{n}_{p\sigma} = \frac{1}{K} \sum_{b=1}^{K}
  \bra{\psi^{(b)}} \hat{a}^{\dagger}_{p\sigma} \hat{a}_{p\sigma}
  \ket{\psi^{(b)}} ,
  \label{eq:occ}
\end{equation}
then fed back into step~3. Iterating until $\bar{n}_{p\sigma}$ and
$\min_{b} E^{(b)}$ stabilize progressively concentrates the subspace on the
dominant configurations. Since the diagonalization cost scales polynomially
in $|\mathcal{S}_{\mathrm{proj}}|$, the procedure remains tractable provided
the number of selected unique bitstrings does not grow exponentially with system
size. In practice the accuracy is governed by how much of the symmetry sector
$\mathcal{S}_{\mathrm{proj}}$ actually covers: when it spans the sector the
energy matches FCI to numerical precision, and restricting the per-batch sample
budget degrades the result systematically~\cite{ksv2026, critical_lims_sqd_2025}.

\end{enumerate}
\subsection{HCI-HSQD}
Heat-Bath Configuration Interaction - Half Sample-Based Quantum
Diagonalization was introduced in Ref.~\cite{McFarthing2026_hcihsqd}.
HCI-HSQD retains the sample-then-diagonalize structure of Sec.~\ref{sub-sec: Theory Sub-Heading 2},
but samples only the $\alpha$ register~\cite{McFarthing2026_hcihsqd, Yoshida2026_dociqsci}
and replaces the fixed-subspace Davidson step with heat-bath selected CI, evaluated
using \textsc{Dice}~\cite{Holmes2016_HCI, Sharma2017_SHCI, Smith2017_HCISCF, Li2018_FastSHCI}
through its PySCF interface~\cite{sun2017pyscf}.
The procedure is as follows.

\begin{enumerate}

\item \textbf{Half-LUCJ ansatz on $M/2$ qubits.} We use the half-LUCJ ansatz quantum
circuit, which halves the register relative
to the full UCJ ansatz by restricting the jastrow operators within one spin spector and sampling configurations from that spin only ~\cite{McFarthing2026_hcihsqd}.
The state $\ket{\psi_{\mathrm{init}}} = \hat{U}_{\mathrm{HSQD}}(\theta) \ket{\psi^{\alpha}_{\mathrm{HF}}}$
is prepared on $M/2$ qubits using the same CCSD-derived
parameters~\cite{lucj_ansatz, Motta2021_lowrank_quCCSD}. We then measure the circuit
and obtain bit-strings that correspond only to the $\alpha$ spin sector. Noisy samples
that violate the electron number count are discarded, and the unique valid bit-strings
are stored as $\mathcal{P}_{\mathrm{unique}}$.

\item \textbf{Tensor-product reconstruction.} The determinant space is rebuilt from the
unique recovered $\alpha$ strings $\mathcal{P}_{\mathrm{unique}}$ as
\begin{equation}
  \mathcal{S}_{\mathrm{proj}} = \mathcal{P}_{\mathrm{unique}} \otimes
  \mathcal{P}_{\mathrm{unique}} ,
  \label{eq:hsqd_sproj}
\end{equation}
which yields bit-strings with the correct Hamming weight and length. The construction
mirrors the self tensor product of Eq.~\eqref{eq:sproj}, but here the two factors are
identical by
construction rather than by post-selection~\cite{sqd_first_paper, McFarthing2026_hcihsqd}.

\item \textbf{Heat-bath selected CI on the reconstructed space.} Instead of diagonalizing
$\mathcal{S}_{\mathrm{proj}}$ directly, we use it to seed the SHCI variational space,
setting $\mathcal{V}^{(0)} = \mathcal{S}_{\mathrm{proj}}$ in place of the single
Hartree--Fock determinant used by
default~\cite{Holmes2016_HCI, Sharma2017_SHCI, Li2018_FastSHCI}. For large active spaces, where the full tensor product in Eq.~\eqref{eq:hsqd_sproj} becomes prohibitively large, $\mathcal{S}_{\mathrm{proj}}$ is truncated to a fixed budget of the sampled strings closest to the Hartree--Fock string. Given the current
variational solution $\ket{\Psi^{(k)}} = \sum_{i \in \mathcal{V}^{(k)}} c^{(k)}_{i}\ket{D_i}$,
the space is expanded by promoting every connected determinant that exceeds the heat-bath
threshold:
\begin{equation}
  \mathcal{V}^{(k+1)} = \mathcal{V}^{(k)}
  \bigcup_{i \in \mathcal{V}^{(k)}}
  \Bigl\{ \ket{D_a} \notin \mathcal{V}^{(k)} \; : \;
  |H_{ai}| > \varepsilon_{\mathrm{var}} / \bigl| c^{(k)}_{i} \bigr| \Bigr\}.
  \label{eq:hci}
\end{equation}
This is equivalent to $|H_{ai} c^{(k)}_{i}| > \varepsilon_{\mathrm{var}}$ for at least one
$i$~\cite{Holmes2016_HCI, Smith2017_HCISCF}, but it is written with the threshold
rescaled per reference to reflect the form actually evaluated. The variational threshold $\varepsilon_{\mathrm{var}}$ is the single parameter that controls the size of the variational space. The product $|H_{ai} c^{(k)}_{i}|$ estimates how strongly an outside determinant $\ket{D_a}$ would contribute to the wave function through its coupling to a determinant already in the space, so only determinants whose contribution exceeds $\varepsilon_{\mathrm{var}}$ are added. Lowering $\varepsilon_{\mathrm{var}}$ admits more weakly coupled determinants, which increases both the accuracy and the cost, and in the limit $\varepsilon_{\mathrm{var}} \to 0$ the procedure recovers full CI. In each round, we
diagonalize the Hamiltonian projected onto the currently selected set:
\begin{equation}
  \hat{H}_{\mathcal{V}^{(k)}} = \hat{P}_{\mathcal{V}^{(k)}} \, \hat{H} \,
  \hat{P}_{\mathcal{V}^{(k)}}, \qquad
  \hat{P}_{\mathcal{V}^{(k)}} = \sum_{i \in \mathcal{V}^{(k)}}
  \ket{D_i}\bra{D_i},
  \label{eq:hci_proj}
\end{equation}
ensuring that only the $|\mathcal{V}^{(k)}| \times |\mathcal{V}^{(k)}|$ block of matrix
elements $H_{ij}$ with $i,j \in \mathcal{V}^{(k)}$ is constructed. This sparse matrix is
diagonalized using the Davidson method~\cite{Davidson1975_iterative_eigensolver} to obtain
the variational energy $E^{(k)}_{\mathrm{var}}$, an upper bound to the true
ground-state energy~\cite{MacDonald1933_RayleighRitz}, and the updated coefficients
$c^{(k)}_{i}$. These coefficients, in turn, set the per-reference thresholds
$\varepsilon_{\mathrm{var}}/|c^{(k)}_{i}|$ used in the next selection round. This iterative process
is repeated, terminating once $|E^{(k)}_{\mathrm{var}} - E^{(k-1)}_{\mathrm{var}}|$ falls
below a prescribed tolerance, at which point the converged $E^{\mathrm{var}}$ and
$\ket{\Psi}$ are returned as the final energy and ground-state wave function. Seeding the
selection with quantum-sampled determinants rather than the Hartree--Fock reference is what
addresses the compactness bottleneck of purely sampled
subspaces~\cite{critical_lims_sqd_2025, McFarthing2026_hcihsqd}.
\item \textbf{Second-order perturbative correction.} At large active-space sizes the converged variational space captures only a small fraction of the correlation energy, and the remainder is recovered by an Epstein-Nesbet second-order perturbative (PT2) correction ~\cite{Epstein1926, Nesbet1955}over the determinants connected to, but not contained in, the final variational space 
\begin{equation}
  E_{\mathrm{PT2}} = \sum_{a \notin \mathcal{V}}
  \frac{\bigl( \sum_{i \in \mathcal{V}} H_{ai} c_{i} \bigr)^{2}}
       {E_{\mathrm{var}} - H_{aa}},
  \qquad
  E_{\mathrm{total}} = E_{\mathrm{var}} + E_{\mathrm{PT2}} .
  \label{eq:hci_pt2}
\end{equation}
The number of external determinants $a$ in Eq.~\eqref{eq:hci_pt2} grows very rapidly with the size of the active space, so the sum cannot be evaluated exactly. It is instead evaluated semistochastically~\cite{Sharma2017_SHCI, Li2018_FastSHCI}, using two thresholds on the size of each contribution $|H_{ai} c_{i}|$, in the same way that $\varepsilon_{\mathrm{var}}$ is used in Eq.~\eqref{eq:hci}. Contributions larger than the deterministic threshold $\varepsilon_{\mathrm{det}}$ are few in number but carry most of the correction, so they are summed exactly. Contributions between $\varepsilon_{\mathrm{det}}$ and the smaller stochastic threshold $\varepsilon_{\mathrm{stoch}}$ are too many to sum one by one, so their total is estimated by random sampling of the variational wave function. The sampling is repeated until the statistical error falls below a chosen target. Contributions smaller than $\varepsilon_{\mathrm{stoch}}$ are neglected. A smaller $\varepsilon_{\mathrm{stoch}}$ therefore gives a more complete correction at a higher cost, and the thresholds satisfy $\varepsilon_{\mathrm{stoch}} < \varepsilon_{\mathrm{det}} < \varepsilon_{\mathrm{var}}$.

The water trimer and the STO-3G carmofur--M\textsuperscript{pro} calculations were performed without the PT2 correction. These systems have at most 16 spatial orbitals per dimer, so the variational space already covers nearly the entire determinant space, and the PT2 correction would be negligible. The 6-31G calculation, with 52 spatial orbitals per dimer, was performed with the PT2 correction. Here the full determinant space exceeds $10^{29}$ determinants, while the variational space reaches only about $10^{6}$, so a large part of the correlation energy lies outside the variational space and must be recovered by PT2 to reach chemical accuracy.
\end{enumerate}

\subsection{Quantum Virtual Machine}
\label{subsec:qvm}

A Quantum Virtual Machine (QVM) provides a hardware-aware physical-resource
virtualization layer between logical quantum circuits and a physical quantum
processor. Instead of treating the QPU as a single indivisible execution
resource or relying on fixed physical-qubit assignments, the QVM represents
the processor as a set of dynamically allocatable execution regions derived
from its physical resources. The underlying processor is represented by a
coupling graph, where vertices denote physical qubits and edges denote
available two-qubit interactions. This abstraction separates the logical
description of a circuit from its physical execution location, allowing
independent workloads to be dynamically assigned to suitable portions of the
same QPU while preserving their logical execution context~\cite{DynQ}.
The region discovery, region-quality score, allocation fitness, and multi-region composition described in this section are adapted from  DynQ~\cite{DynQ}. The contributions of this work are the fragment-batching layer that feeds the FMO circuits to the allocator and the BRB temporal reuse of regions.

\begin{figure}[H]
\centering
\includegraphics[width=0.7\linewidth]{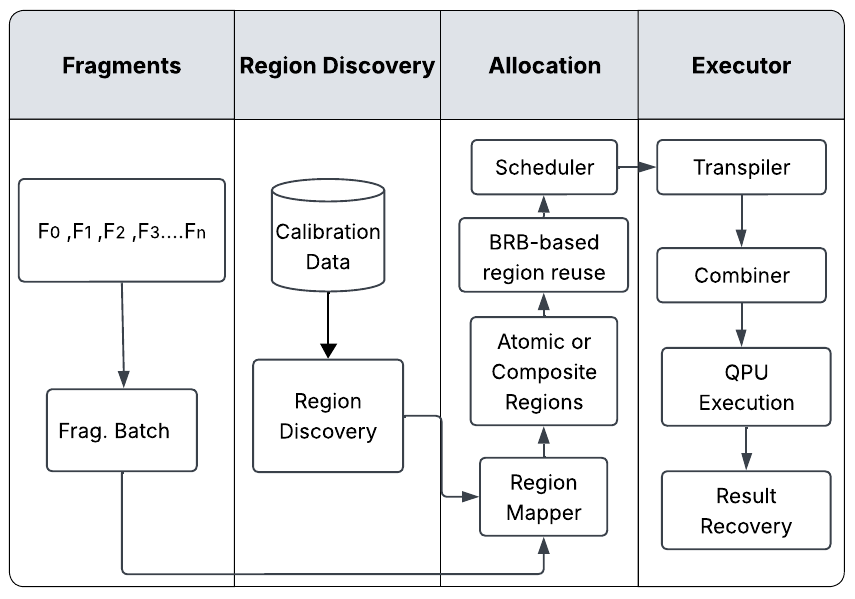}
\caption{Workflow of the QVM for executing fragment circuits on a shared QPU. \emph{Fragments}: the monomer and dimer circuits $F_0,\dots,F_n$ generated from the FMO Hamiltonians are collected into a fragment batch. \emph{Region Discovery}: calibration data of the target backend are converted into a quality-weighted coupling graph, which is partitioned by community detection into a pool of atomic execution regions. \emph{Allocation}: the Region Mapper matches each circuit in the batch to the region pool, placing it on an atomic region or, if its qubit requirement exceeds every free atomic region, on a composed region. Circuits that cannot be placed because all suitable regions are occupied reuse a region through BRB (barrier--reset--barrier) temporal reuse, and the Scheduler fixes the resulting execution order. \emph{Executor}: the Transpiler compiles each circuit onto the coupling map of its assigned region, the Combiner merges the compiled circuits into a single job for the QPU, and Result Recovery maps the measured bitstrings back to the originating fragments.}
\label{fig:qvm_workflow}
\end{figure}

\subsubsection{Hardware-Aware Region Discovery}
\label{subsubsec:region_discovery}
The \textit{Region Discovery} stage in Fig.~\ref{fig:qvm_workflow} constructs a hardware-aware representation of the target quantum processor from its physical connectivity and calibration information. The processor is represented as a quality-weighted coupling graph
\begin{equation}
G=(V,E,w),
\label{eq:qvm_graph}
\end{equation}
where $V$ denotes the physical qubits, $E$ denotes the available two-qubit couplings, and $w$ assigns a quality weight to each coupler. Following DynQ~\cite{DynQ}, the weight of the coupler between qubits $i$ and $j$ is the inverse of its calibrated two-qubit gate error,
\begin{equation}
w_{ij}=\frac{1}{\epsilon_{ij}+\epsilon_0},
\label{eq:edge_weight}
\end{equation}
where $\epsilon_{ij}$ is the two-qubit gate error rate reported in the backend calibration data and $\epsilon_0=10^{-6}$ is a small constant that avoids division by zero. We set $w_{ij}=0$ if qubits $i$ and $j$ are not coupled. Couplers with low error therefore receive large weights, so that the graph captures both the connectivity and the non-uniform quality of the processor.

The weighted graph is partitioned using community detection to identify groups of physical qubits with strong internal connectivity. The QVM uses the Louvain algorithm~\cite{Louvain,DynQ} to maximize the weighted modularity
\begin{equation}
Q_{\mathrm{mod}} =
\frac{1}{2W}
\sum_{i,j}
\left(
w_{ij}-\frac{s_i s_j}{2W}
\right)
\delta(c_i,c_j),
\label{eq:qvm_modularity}
\end{equation}
where $c_i$ is the community assigned to qubit $i$, $s_i=\sum_j w_{ij}$ is the weighted degree of qubit $i$, and $W=\frac{1}{2}\sum_{i,j}w_{ij}$ is the total edge weight of the graph. The function $\delta(c_i,c_j)$ equals one when qubits $i$ and $j$ belong to the same community and zero otherwise. The term $s_i s_j/2W$ is the weight expected between $i$ and $j$ in a random graph that preserves the weighted degrees, so maximizing $Q_{\mathrm{mod}}$ favors communities whose internal connections are stronger than expected by chance~\cite{Louvain,DynQ}. Because $w_{ij}$ is large for low-error couplers, community boundaries tend to fall along the noisiest couplers. We use the standard resolution parameter $\gamma=1$ and fix the Louvain random seed to 42.

The resulting communities form the \textit{atomic execution regions} shown as the output of the Region Discovery block in Fig.~\ref{fig:qvm_workflow}. Each region $R_i\subseteq V$ represents a group of physical qubits that can be exposed to the allocation stage as an independent execution resource. Discovery is performed offline once per calibration cycle, and the resulting regions form the hardware-aware resource pool from which suitable physical resources are selected for incoming circuits~\cite{DynQ}.

\subsubsection{Region Selection and Allocation}
\label{subsubsec:region_allocation}

Following region discovery, the \textit{Region Mapper} in
Fig.~\ref{fig:qvm_workflow} evaluates the available atomic regions for an
incoming circuit according to its resource requirements and the properties
of the candidate regions. For a circuit requiring $n$ logical qubits, a
candidate region $R$, with qubit set $V_R$ and coupler set $E_R$, is
feasible only when $|R|\geq n$, where $|R|$ denotes the number of physical
qubits in the region.

Each candidate region is characterized by four quality components~\cite{DynQ}.
The connectivity score is the edge density of the region,
\begin{equation}
S_{\mathrm{conn}}(R)=\frac{2|E_R|}{|V_R|\left(|V_R|-1\right)}.
\label{eq:qvm_sconn}
\end{equation}
The two-qubit gate quality and readout quality are
\begin{equation}
S_{\mathrm{gate}}(R)=\max\!\left(0,\,1-\alpha_{\mathrm{gate}}\,\bar{\epsilon}_{\mathrm{gate}}\right),
\qquad
S_{\mathrm{ro}}(R)=\max\!\left(0,\,1-\alpha_{\mathrm{ro}}\,\bar{\epsilon}_{\mathrm{ro}}\right),
\label{eq:qvm_sgate_sro}
\end{equation}
where $\bar{\epsilon}_{\mathrm{gate}}$ is the mean two-qubit gate error over
the couplers in $E_R$, $\bar{\epsilon}_{\mathrm{ro}}$ is the mean readout
error over the qubits in $V_R$, and the scale factors
$\alpha_{\mathrm{gate}}=100$ and $\alpha_{\mathrm{ro}}=10$ map typical error
rates onto the interval $[0,1]$. The internal uniformity penalizes variation
of the gate error inside the region,
\begin{equation}
S_{\mathrm{unif}}(R)=\max\!\left(0,\,1-\frac{\sigma_{\epsilon}}{\mu_{\epsilon}}\right),
\label{eq:qvm_sunif}
\end{equation}
where $\mu_{\epsilon}$ and $\sigma_{\epsilon}$ are the mean and standard
deviation of the gate errors in $E_R$.

These components are combined into the composite region-quality score
\begin{equation}
Q(R)=
\beta_{\mathrm{conn}}S_{\mathrm{conn}}(R)
+\beta_{\mathrm{gate}}S_{\mathrm{gate}}(R)
+\beta_{\mathrm{ro}}S_{\mathrm{ro}}(R)
+\beta_{\mathrm{unif}}S_{\mathrm{unif}}(R),
\label{eq:qvm_quality}
\end{equation}
with $\beta_{\mathrm{conn}}=0.25$, $\beta_{\mathrm{gate}}=0.35$,
$\beta_{\mathrm{ro}}=0.20$, and $\beta_{\mathrm{unif}}=0.20$. DynQ fixes the
connectivity weight to one and uses the remaining coefficients to control
the relative importance of gate quality, readout quality, and
uniformity~\cite{DynQ}; we instead choose weights that sum to one, so that,
among the quality components of $Q(R)$, gate quality receives the largest
weight, since two-qubit gates dominate the fragment circuits.

Among feasible regions, the QVM ranks candidates using the online
allocation fitness function
\begin{equation}
F(R,n)
=
w_{\mathrm{size}}S_{\mathrm{size}}(R,n)
+w_{\mathrm{conn}}S_{\mathrm{conn}}(R)
+w_{Q}\,Q(R),
\label{eq:qvm_fitness}
\end{equation}
where $S_{\mathrm{size}}(R,n)$ measures the compatibility between the
region size and the circuit requirement, and $w_{\mathrm{size}}=0.2$, $w_{\mathrm{conn}}=0.4$, and $w_{Q}=0.4$ weight
size match, connectivity, and aggregate region quality. Since
$S_{\mathrm{conn}}$ also enters $Q(R)$, its effective weight in $F$ is 0.5,
so allocation favours compact, well-connected regions.
DynQ multiplies
the size score with the connectivity and quality terms~\cite{DynQ}; we use
an additive form and exclude undersized regions through the feasibility
condition above. The available region with the highest fitness is selected
and marked as occupied for the corresponding circuit execution.

The size score is defined as~\cite{DynQ}
\begin{equation}
S_{\mathrm{size}}(R,n)=
\begin{cases}
0, & |R| < n,\\
1, & |R| = n,\\
\exp\left[-0.5\dfrac{|R|-n}{n}\right], & |R| > n.
\end{cases}
\label{eq:qvm_size_score}
\end{equation}

Thus, an undersized region is infeasible, an exact-fit region receives the
maximum score, and 
an oversized region is progressively penalized. Because
Eq.~\eqref{eq:qvm_fitness} is additive, this penalty is bounded by
$w_{\mathrm{size}}$, so a larger region of higher quality can still be
selected.

When no single free atomic region can accommodate a circuit, the QVM forms a
composed region by the bridge-aware greedy expansion of DynQ~\cite{DynQ}.
Starting from the largest free region, it repeatedly attaches the adjacent
free region $R$ that maximizes
\begin{equation}
  s(R) = 0.4\,Q(R) + 0.4\,S_{\mathrm{conn}}\!\left(G[V \cup V_R]\right)
       + 0.2\,S_{\mathrm{bridge}}(E_B),
  \label{eq:compose}
\end{equation}
until the footprint holds at least $n$ qubits. Here $V$ is the current
footprint, $V_R$ the qubit set of $R$, and $E_B$ the set of couplers joining
$R$ to $V$. The bridge score applies the gate-quality score of
Eq.~\eqref{eq:qvm_sgate_sro} to these couplers,
\begin{equation}
  S_{\mathrm{bridge}}(E_B) = \max\!\left(0,\;
  1 - \frac{\alpha_{\mathrm{gate}}}{|E_B|}
  \sum_{(i,j)\in E_B} \varepsilon_{ij}\right),
  \label{eq:sbridge}
\end{equation}
so that regions joined through low-error couplers are preferred. Because
community boundaries tend to fall along the noisiest couplers, this term
guards against composed regions held together by weak links. The composed
region is assigned the lowest quality score among its constituent regions,
so that it is never rated above its weakest part, and it is released as a
whole once its circuit completes. Composition thereby extends region-based
allocation to circuits whose qubit requirement exceeds every free atomic
region.

When multiple circuits compete for a limited number of available regions,
the QVM additionally reuses physical regions across successive execution
intervals.
In our implementation, a BRB sequence separates
successive circuit occupancies of the same physical region. The barriers
establish execution boundaries, while the reset operation returns the
reused qubits to a known initial state. Thus, spatial allocation through
atomic or composed regions and BRB-based temporal reuse provide
complementary mechanisms for increasing physical-resource utilization.
\subsubsection{Region-Aware Circuit Execution}
\label{subsubsec:region_execution}

Following region selection and allocation, the \textit{Executor} stage in
Fig.~\ref{fig:qvm_workflow} prepares each circuit using only the physical
resources assigned to its region. The \textit{Transpiler} receives the
coupling graph induced by the selected region together with an initial
layout restricted to its physical qubits. This subgraph-restricted
compilation confines mapping, routing, and scheduling operations to the
allocated resources, preventing the compiler from using qubits outside the
selected region~\cite{DynQ,SABRE}.

The resulting region-level circuits are passed to the \textit{Combiner},
which embeds the independently compiled circuits into a common physical
execution context while preserving their region-specific mappings. The
combined workload is then submitted to the target QPU for execution.
The QVM retains the mapping metadata between region-local and
backend-global physical qubits throughout this process, allowing each
circuit to remain logically independent of its physical execution
location~\cite{DynQ}.

Following QPU execution, the \textit{Result Recovery} stage associates the
measured physical outcomes with their originating circuits and restores
their corresponding logical ordering using the retained mapping metadata.
This is achieved through the inverse of the region-to-backend mapping,
allowing the resulting circuit-level measurement distributions to be
returned to the higher-level quantum workflow without exposing the
physical-resource management performed by the QVM~\cite{DynQ}. All circuits in a combined job are executed with the same number of shots.
\section{Methodology}

\label{sec:methodology}
In this section we introduce  a resource efficient workflow used for quantum
simulation of large molecular systems. These methods combine
fragment molecular orbital (FMO) decomposition with sample-based quantum
diagonalization (SQD) and its half-qubit variant (HCI-HSQD) to find the correlated energy of each fragment on quantum hardware while using the QPU efficiently, the
overall system energy is assembled classically through the FMO expansion.

\subsection{Fragmentation}
\label{sub-sec: Fragmentation and active space}
The target molecule is first partitioned into fragments using the
\textsc{FragIt} software and the input is passed to \textsc{OpenFMO}, which
carries out the two-body FMO (FMO2) procedure and produces the embedded
monomer and dimer active-space Hamiltonian integrals for each fragment. We
choose the active space by taking an equal number of the highest occupied
and lowest unoccupied orbitals around the HOMO--LUMO gap, keep the remaining
occupied orbitals frozen and discard the remaining virtual orbitals. 

\begin{figure}[H]
    \centering
    \includegraphics[width=0.9\linewidth]{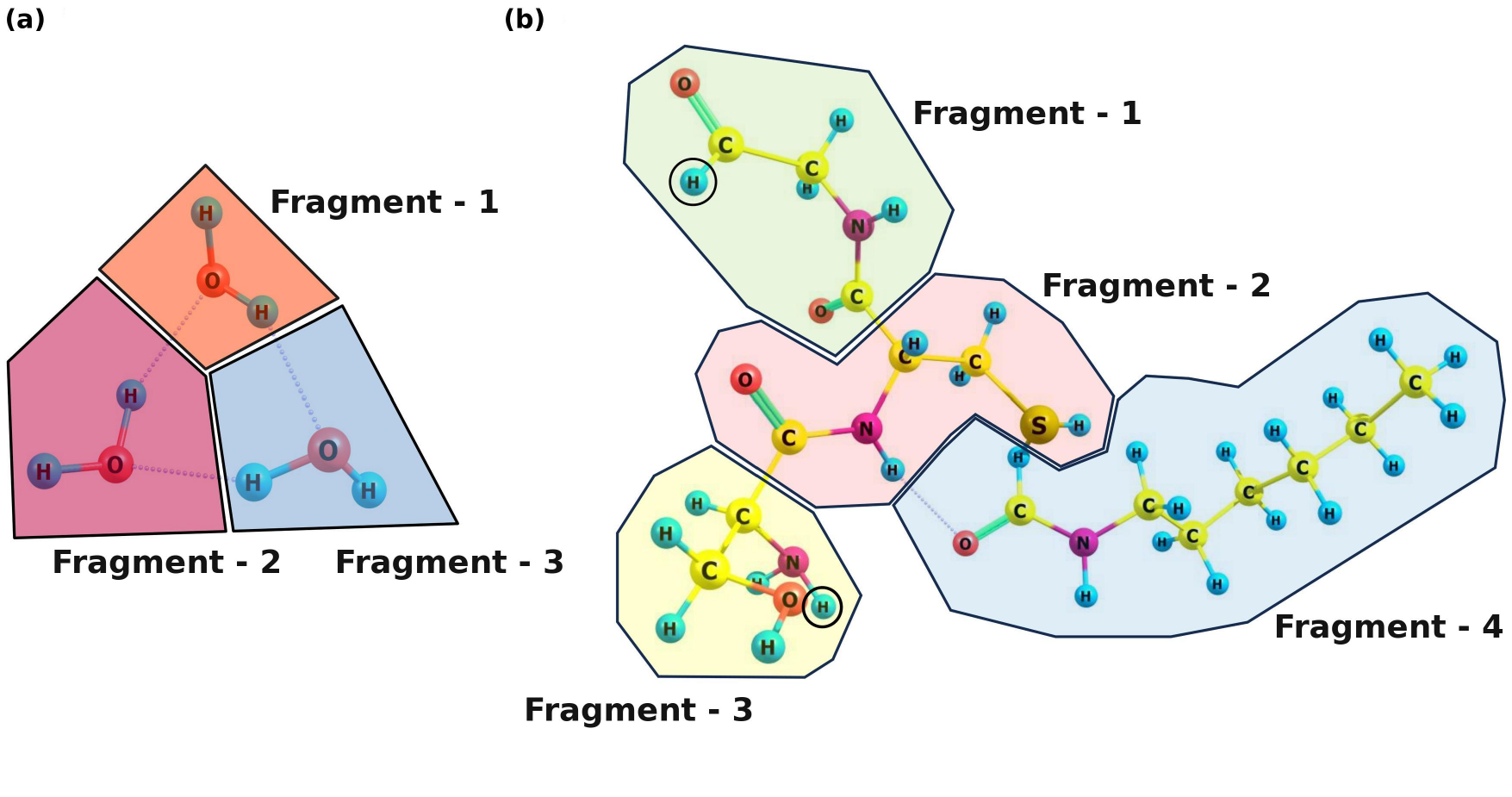}
    \caption{The fragmentation scheme for the two molecules studied in this paper. The water trimer (Fig a.) is cut into 3 fragments while the carmofur--M\textsuperscript{pro} complex model (Fig b.) is cut into 4 fragments.}
    \label{fig:Fragment-scheme}
\end{figure}

In this work we study two systems, shown in Fig.~\ref{fig:Fragment-scheme}.
The first is the water trimer, $(\mathrm{H_2O})_3$, in which three water
molecules interact through a network of hydrogen bonds. It serves as a
well-characterized benchmark for which an exact FMO-FCI reference energy is
available, allowing us to quantify the accuracy of the quantum solvers
directly. Each water molecule is treated as one fragment, and the full orbital
space of each fragment is used, giving (10e,\,7o) per monomer and
(20e,\,14o) per dimer in the STO-3G basis. The second system is the SARS-CoV-2
main protease (M\textsuperscript{pro}), an established antiviral drug target,
in complex with the covalent inhibitor carmofur. On binding, the Cys145
thiolate attacks the carbamoyl carbon of carmofur, releasing the
5-fluorouracil moiety and leaving the hexylcarbamoyl tail covalently bound to
Cys145 through a 1.8~\AA{} C--S bond~\cite{Jin2020_Mpro}. We therefore
model this covalently bound tail together with the residues Ser144, Cys145 and
Gly146 that surround the attachment site. This region is partitioned with
\textsc{FragIt} into four fragments, one per residue and one for the ligand,
with boundaries placed along single bonds. More refined partitioning schemes
exist~\cite{dmet_sqd, Bowling2025_ProteinLigand_FragQC}, but this
fragmentation is sufficient to benchmark FMO-QVM-HCI-HSQD on a realistic
protein--ligand system. For each fragment, we choose the active space by
taking an equal number of the highest occupied and lowest unoccupied orbitals
around the HOMO--LUMO gap, keeping the remaining occupied orbitals frozen and
discarding the remaining virtual orbitals. In the STO-3G basis we use
(8e,\,8o) per monomer and (16e,\,16o) per dimer. In the 6-31G basis we use
(26e,\,26o) per monomer and (52e,\,52o) per dimer, as shown in
Fig.~\ref{fig: Carmofur_active}. These integrals define the effective
electronic structure problem that is subsequently handed to the quantum
solver.
\\

The fragmentation scheme for both systems is described in Fig \ref{fig:Fragment-scheme}. The data files containing the fragment Hamiltonian integrals are used by our quantum algorithms  to obtain their fragment energies, which is then assembled using the FMO formula described in Eqn. 5-11.

Based on Eq.\ref{eq:num-hamiltonians}, \textbf{for the two systems studied here gives $N_{\mathrm{H}}=6$ for the water
trimer ($N=3$: three monomers and three dimers) and $N_{\mathrm{H}}=10$ for the
carmofur--M\textsuperscript{pro} fragment model ($N=4$: four monomers and six dimers).}

\subsection{Parallelization with QVM Integration}

\textbf{Sequential run of Monomers and Dimers without QVM (FMO--SQD/HCI-HSQD):}
Once \textsc{OpenFMO} has generated the monomer and dimer Hamiltonian
integrals together with their SCC-converged reference energies, each fragment
Hamiltonian is passed to the quantum solver for evaluation of its ground-state
energy by the SQD and HCI-HSQD algorithms. In the absence of the QVM layer,
the fragment Hamiltonians to be run on hardware are dispatched to the quantum
processor sequentially, each as an independent job. This serial execution is
inefficient in two respects. First, the total quantum-processor time grows
with the number of fragments, since every job incurs its own fixed overhead.
Each job waits separately in the queue, its circuit is compiled and its
control pulses are loaded onto the device, and the results are transferred
back to the classical host. On top of this, the qubit reset and readout
between shots are paid for every circuit. Because FMO2 requires a separate
circuit for every monomer and dimer, these costs accumulate rapidly as the
number of fragments increases. Second, each fragment circuit occupies only a
small number of qubits, set by the size of its active space, so a single
fragment leaves the majority of the device's qubits idle. Executing the
fragments one at a time therefore leaves a large fraction of the available
hardware unused throughout the calculation, an inefficient use of the quantum
resource.
\\

\textbf{QVM-assisted execution of monomer and dimer Hamiltonians
(FMO--QVM--SQD/HCI-HSQD):}
To perform hardware-aware resource allocation, the generated fragment circuits are submitted to the QVM. Fragments are assigned to available atomic execution regions, with compatible regions composed together when a fragment's qubit requirement exceeds a single region's capacity. The hardware mapping for the carmofur--M\textsuperscript{pro} HCI-HSQD workload on \texttt{ibm\_marrakesh} is shown in Fig.~\ref{fig:qvm_hsqd_allocation} and the corresponding FMO-SQD allocation is given in Appendix~\ref{appendixA}. 
\begin{figure}[H]
\centering
\includegraphics[width=1.0\linewidth]{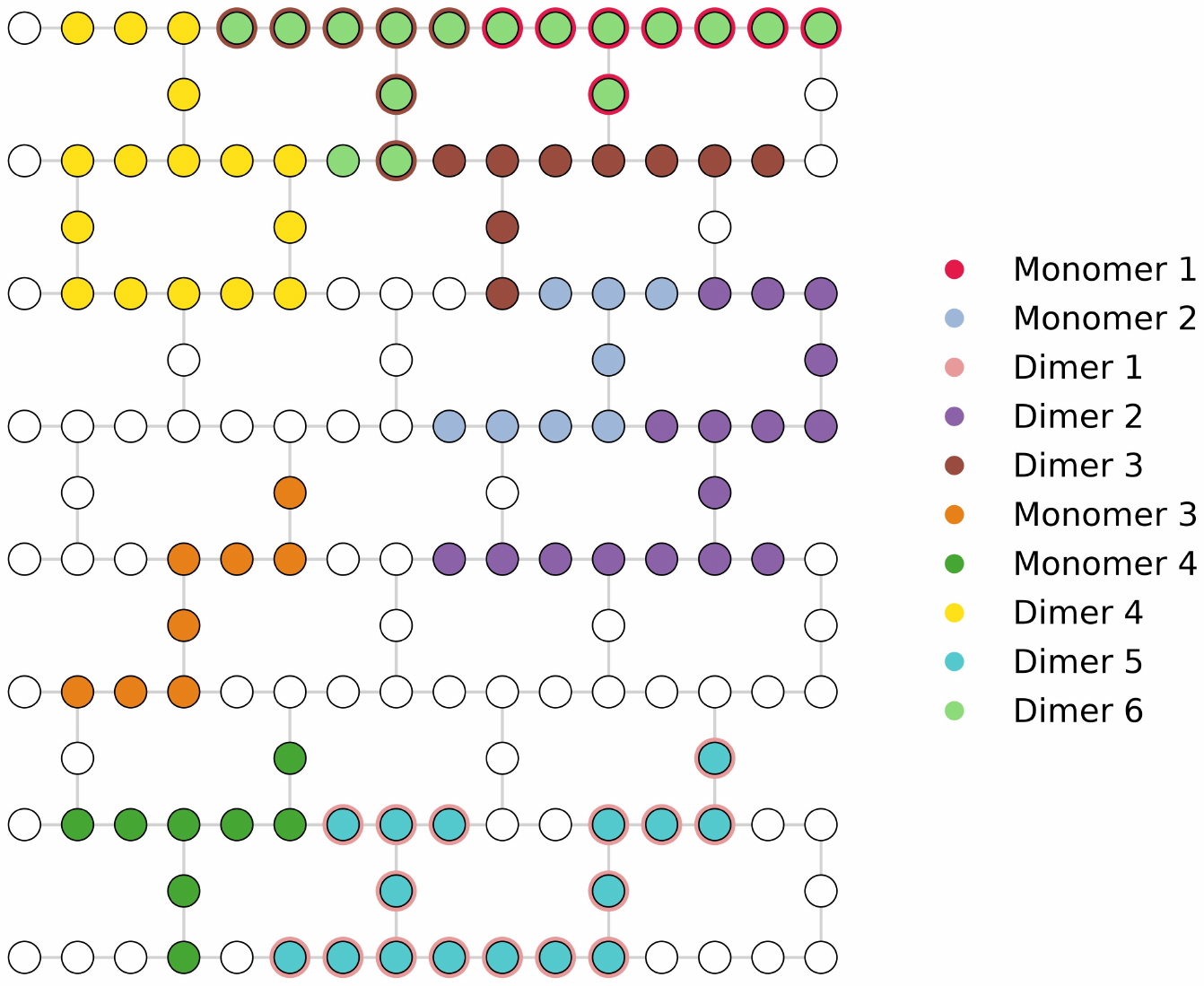}
\caption{QVM-based physical allocation of the carmofur--M\textsuperscript{pro} FMO-QVM-HCI-HSQD fragments (STO-3G) on \texttt{ibm\_marrakesh}. Fill colour marks the first fragment placed on each qubit; an outer ring marks a later fragment that reuses the qubit through BRB.}
\label{fig:qvm_hsqd_allocation}
\end{figure}
\begin{figure}
    \centering
    \makebox[\linewidth][c]{\includegraphics[width=1.1\linewidth]{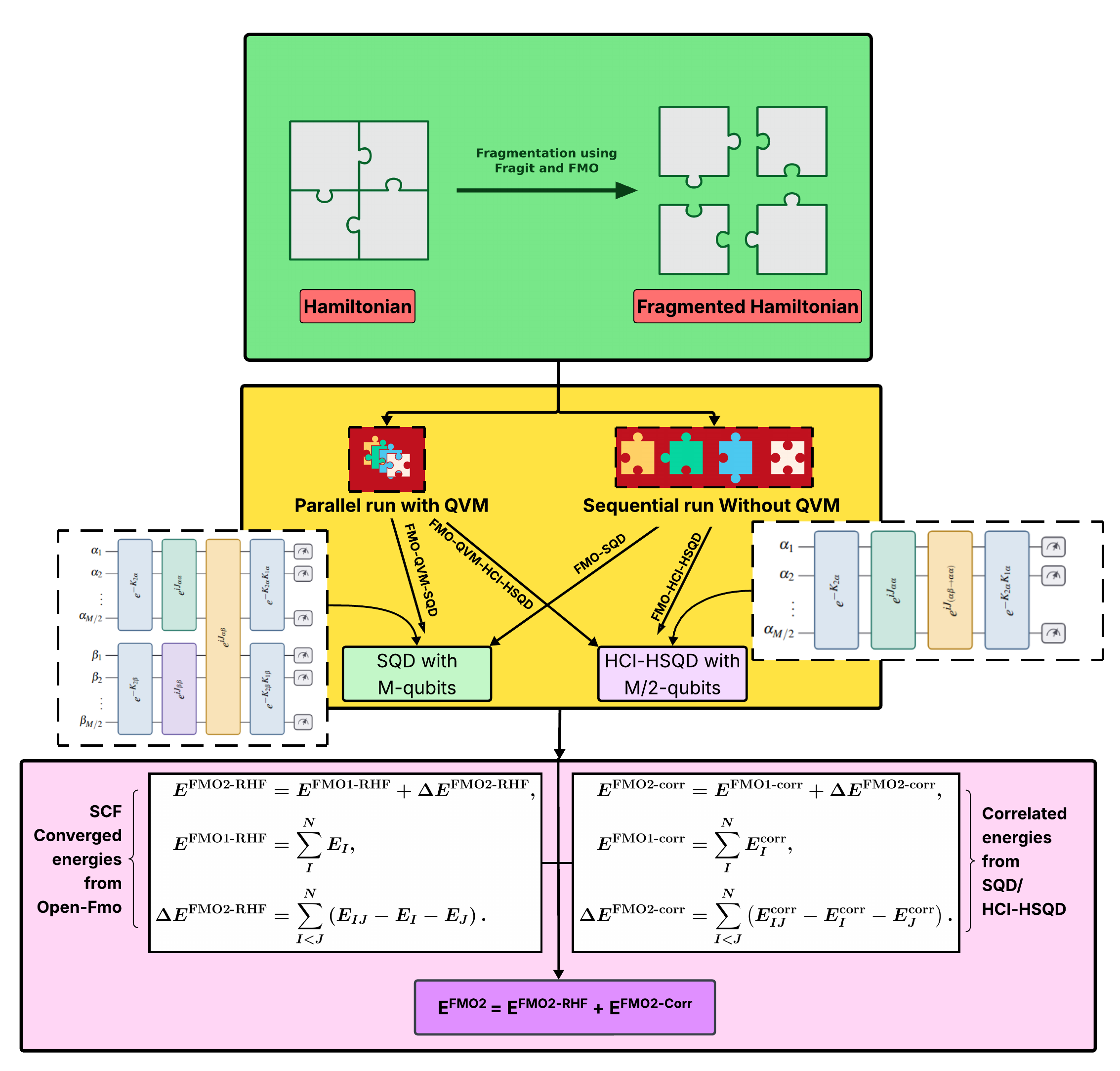}}
    \caption{Complete workflow of the FMO-based quantum pipeline. Fragment Hamiltonian integrals are passed to the different quantum solvers, with or without quantum virtual machines (QVMs), and the resulting monomer and dimer energies are combined using the FMO expansion to obtain the total ground-state energy of the molecule.}
    \label{fig:fmo_workflow}
\end{figure}
In addition to spatial allocation, the implementation supports temporal reuse of physical
regions: a region occupied by one fragment circuit can be reused by a later circuit within the same job, separated
by the BRB sequence of Sec. \ref{subsubsec:region_allocation}.
The temporal reuse is explicitly visualized in Fig.~\ref{fig:qvm_hsqd_allocation} via concentric rings on the physical qubit nodes. After execution, the QVM-generated measurement outcomes are collected as bitstring samples for each fragment and passed to the corresponding HCI-HSQD post-processing pipeline.
The FMO-SQD workload further demonstrates the same allocation mechanism for larger fragment requirements, with dimer fragments requiring 32 qubits and monomer fragments requiring 16 qubits. The corresponding physical mapping and detailed fragment-to-region assignments are provided in
Appendix~\ref{appendixA}. For the water trimer, only the three dimer circuits are executed on hardware; for each solver, they run concurrently in a single job on \texttt{ibm\_marrakesh} and the resulting resource comparison is discussed in Sec.~\ref{sub-sec: Energy Comparison}.
\\

In the 6-31G calculation on \texttt{ibm\_boston}, where each monomer requires 26 qubits and each dimer 52 qubits, all ten fragment circuits were combined into a single job. The QVM formed three physically disjoint composed regions: two of 27 and 28 qubits for the monomers and one of 53 qubits for the dimers. Three circuits were placed spatially and the remaining seven through BRB temporal reuse, giving six execution batches. In each of the first two batches, two monomers were executed concurrently with one dimer, while the remaining four dimers were executed one after another in the last four batches, because the device could not host several 52-qubit circuits in parallel on qubits of sufficient quality. The QPU time reported in Fig.~\ref{fig:Carmofur_631-g} therefore reflects six execution batches rather than ten: the monomers add no batches of their own, while the dimers gain no concurrency among themselves.
\section{Results and Discussion}
\label{sec:results_and_discussion}

In this section we present simulation results for the molecules shown in Fig.~\ref{fig:Fragment-scheme}. For the water trimer, we compare all four methods (FMO-SQD, FMO-HCI-HSQD, FMO-QVM-SQD and FMO-QVM-HCI-HSQD). For the carmofur--M\textsuperscript{pro} complex in the STO-3G basis, we compare FMO-QVM-SQD with FMO-QVM-HCI-HSQD. Finally, we apply FMO-QVM-HCI-HSQD to the same complex in the 6-31G basis. The fragmentation and active spaces used for STO-3g basis runs and the 6-31g runs are described in detail in  sub-section \ref{sub-sec: Fragmentation and active space}. The classical references we use for comparison are FMO-FCI, FMO-CASCI and FMO-HCI.

\subsection{Resource-Efficient Quantum Simulation with FMO-QVM-HCI-HSQD}
\label{sub-sec: Results Sub-Heading 1}

\subsubsection{Energy Comparison}
\label{sub-sec: Energy Comparison}
\begin{figure}[H]
    \centering
    \includegraphics[width=1.0\linewidth]{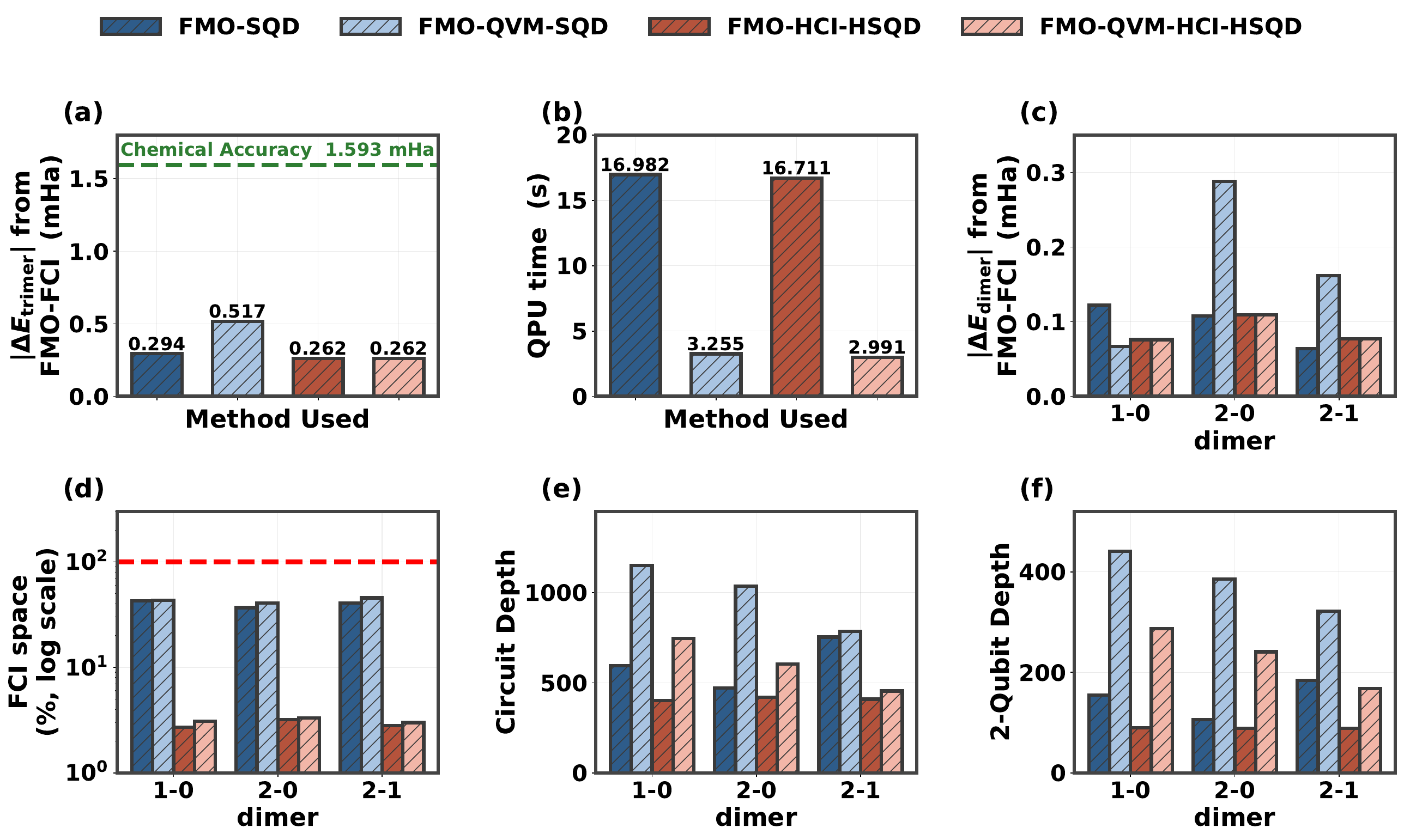}
    \caption{Results for the water trimer using all four methods. The monomers are treated with FCI. The dimers require 28 qubits with SQD and 14 qubits with HCI-HSQD. All calculations use the STO-3G basis with 10{,}000 shots on \texttt{ibm\_marrakesh}. The HCI step in the HCI-HSQD variants uses only the variational expansion, without the PT2 correction. (a)~Absolute deviation of the FMO2 total energy from FMO-FCI. (b)~Total QPU time. (c)~Absolute deviation of each dimer energy from FCI. (d)~Fraction of the symmetry-allowed determinant space accessed by each method for each dimer. (e)~Circuit depth and (f)~two-qubit gate depth of each dimer circuit.}
    \label{fig: Water_FMO_Summary}
\end{figure}
For the water trimer, we assemble the FMO total energy for all four solvers in a hybrid scheme in which the monomer energies are taken from FCI, which is tractable at 14 spin orbitals per fragment, and only the dimer energies come from the solver under test. Figure~\ref{fig: Water_FMO_Summary} summarises the results. Panel~(a) of Figure~\ref{fig: Water_FMO_Summary}  shows the deviation of the assembled total energy from FMO-FCI, and panel~(c) of the same shows the dimer deviations from FMO-FCI. All deviations are within chemical accuracy. The total-energy deviations are {0.294}~mHa for FMO-SQD, {0.517}~mHa for FMO-QVM-SQD, {0.262}~mHa for FMO-HCI-HSQD and {0.262}~mHa for FMO-QVM-HCI-HSQD. Both HCI-HSQD variants give a slightly lower total-energy deviation than the two SQD variants, while accessing only about $3\%$ of the symmetry-allowed determinant space, compared with about $40\%$ for SQD. QVM execution decreases the accuracy when paired with SQD but leaves the accuracy essentially unchanged when paired with HCI-HSQD at this orbital count, so FMO-QVM-HCI-HSQD combines the accuracy of HCI-HSQD with the reduced QPU time of parallel execution.

\begin{figure}[H]
    \centering
    \includegraphics[width=1.0\linewidth]{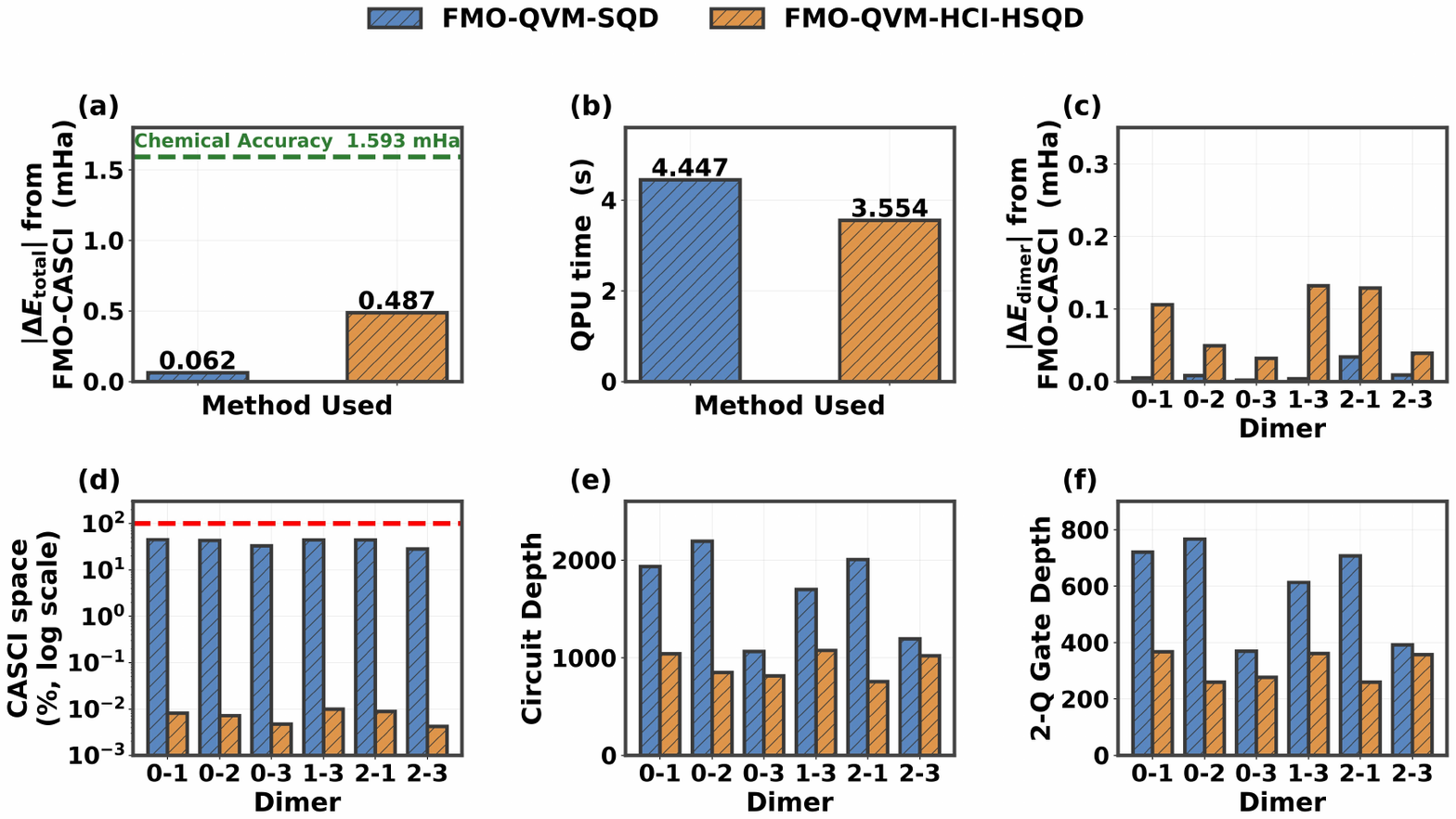}
    \caption{Results for the carmofur--M\textsuperscript{pro}  using the QVM-assisted methods. The monomers are treated with CASCI. The dimers require 32 qubits with SQD and 16 qubits with HCI-HSQD. All calculations use the STO-3G basis with 10{,}000 shots on \texttt{ibm\_marrakesh}. The HCI step in FMO-QVM-HCI-HSQD uses only the variational expansion, without the PT2 correction. (a)~Absolute deviation of the FMO2 total energy from FMO-CASCI. (b)~Total QPU time. (c)~Absolute deviation of each dimer energy from CASCI. (d)~Fraction of the CASCI determinant space accessed by each method for each dimer, on a logarithmic scale. (e)~Circuit depth and (f)~two-qubit gate depth of each dimer circuit.}
    \label{fig:Carmofur_comparison}
\end{figure}

For the carmofur--M\textsuperscript{pro} complex, we compare FMO-QVM-SQD with FMO-QVM-HCI-HSQD. Since the water trimer results show that QVM reduces the QPU time at similar accuracy, a sequential run was not performed. The comparison is shown in Fig.~\ref{fig:Carmofur_comparison}. As for the water trimer, the deviations are computed against FMO-CASCI, using CASCI energies for the monomers and the respective solvers for the dimers. Both methods are well within chemical accuracy, with total-energy deviations of 0.062~mHa for FMO-QVM-SQD and 0.487~mHa for FMO-QVM-HCI-HSQD, and dimer deviations of at most 0.034~mHa and 0.132~mHa, respectively. FMO-QVM-SQD gives a smaller deviation, but across the dimers it accesses  28--45\% of the CASCI determinant space, whereas FMO-QVM-HCI-HSQD accesses only 0.004--0.010\%, a subspace nearly four orders of magnitude smaller. Since our goal is to reach chemical accuracy with minimal resources, we report results at these settings.
\subsubsection{Hardware Resource Comparison}
For the water trimer, we also compare the four algorithms in terms of circuit depth, two-qubit gate depth, QPU time and the quality of the sampled bitstrings. Two comparisons are relevant here: QVM against sequential execution, and SQD against HCI-HSQD. The primary motivations for HCI-HSQD are to halve the number of qubits per dimer and to reduce the circuit depth, thereby decreasing the effect of hardware noise on the sampled bitstrings. Relative to SQD, HCI-HSQD reduces the total circuit depth by 33\% and the two-qubit gate depth by 40\% in sequential execution, and both by 39\% under QVM (Fig.~\ref{fig: Water_FMO_Summary}, panels~e and~f). The sampling quality of all four methods is discussed in detail in Appendix~\ref{app:sampling}.

Parallel execution under QVM reduces the QPU time by roughly a factor of five (Fig.~\ref{fig: Water_FMO_Summary}, panel~b), but it increases the circuit depth relative to sequential execution (Fig.~\ref{fig: Water_FMO_Summary}, panels~e and~f). When a fragment is executed alone, the transpiler can place it on the best-connected, lowest-error qubits of the entire device. Under QVM, the device is shared among all fragments, and each circuit is restricted to its own region. The qubits in that region are not necessarily well connected for the interaction pattern of the circuit, so additional SWAP gates are required to route the two-qubit gates on the heavy-hex lattice. These SWAP gates increase both the total and the two-qubit gate depth, and with them the exposure to hardware noise. The QPU times of sequential FMO-SQD (16.98~s) and FMO-HCI-HSQD (16.71~s) are nearly identical, which shows that the reduction in QPU time arises from parallel execution under QVM rather than from the reduced depth of HCI-HSQD. FMO-QVM-HCI-HSQD combines both advantages. Its QPU time of 2.99~s is $5.7\times$ shorter than that of sequential FMO-SQD, and its total circuit depth is 39\% lower than that of FMO-QVM-SQD, which brings it close to the depth of sequential FMO-SQD.

For the carmofur--M\textsuperscript{pro} complex, panels~(e) and~(f) of Fig.~\ref{fig:Carmofur_comparison} show that HCI-HSQD reduces both the circuit depth and the two-qubit gate depth. This reduces the effect of hardware noise on the sampled bitstrings and therefore improves the quality of the bitstrings used to build the subspace. Panel~(b) compares the QPU time of the two QVM methods. Under QVM, all fragment circuits are executed simultaneously in a single job, so every shot samples all fragments at once and the duration of each shot is set by the deepest circuit in the batch. The QPU time is therefore governed by the number of shots, the fixed per-shot overhead of qubit reset and readout, and the depth of the deepest fragment circuit, rather than by the number of fragments. Because HCI-HSQD shortens the fragment circuits, FMO-QVM-HCI-HSQD requires 3.55~s of QPU time compared with 4.45~s for FMO-QVM-SQD, a reduction of 20\%. This reduction is smaller than the reduction in depth mainly because the per-shot reset and readout overhead does not depend on the circuit depth.

\

\subsection{Experiments upto 52 qubits with FMO-QVM-HCI-HSQD}

\begin{figure}[H]
    \centering
    \includegraphics[width=1.0\linewidth]{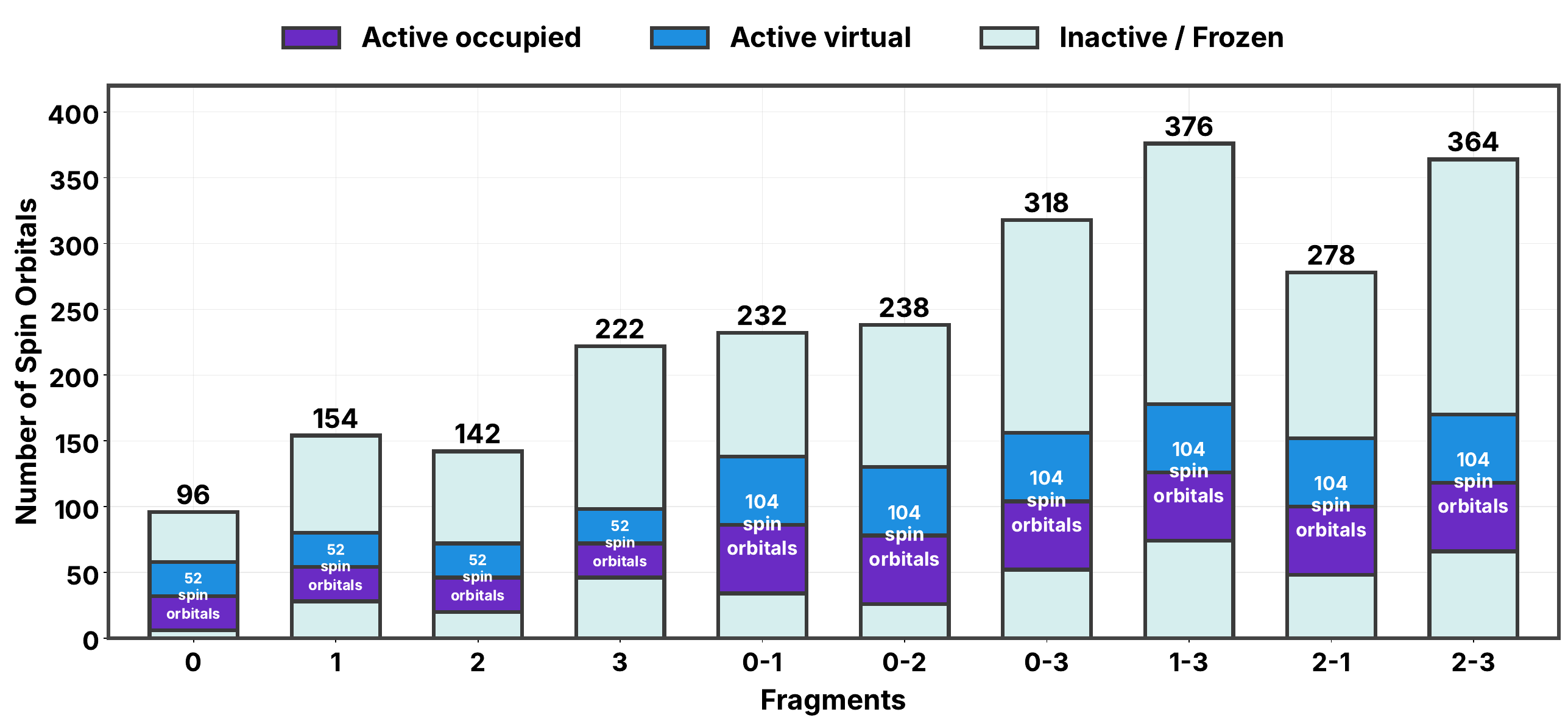}
    \caption{Active space chosen for each fragment in the 6-31G calculation of the carmofur--M\textsuperscript{pro} complex. Each bar shows the spin orbitals of one fragment, divided into active occupied (purple), active virtual (blue) and inactive or frozen (light) orbitals. The number on top of each bar is the total number of spin orbitals of the fragment, and the number inside the bar is the number of active spin orbitals.}
    \label{fig: Carmofur_active}
\end{figure}

In this sub-section we move to a larger active space simulation with FMO-QVM-HCI-HSQD. We study the same carmofur--M\textsuperscript{pro} complex, this time in the 6-31G basis with a much larger active space, described in Fig.~\ref{fig: Carmofur_active}.

\begin{figure}
    \centering
    \includegraphics[width=1.0\linewidth]{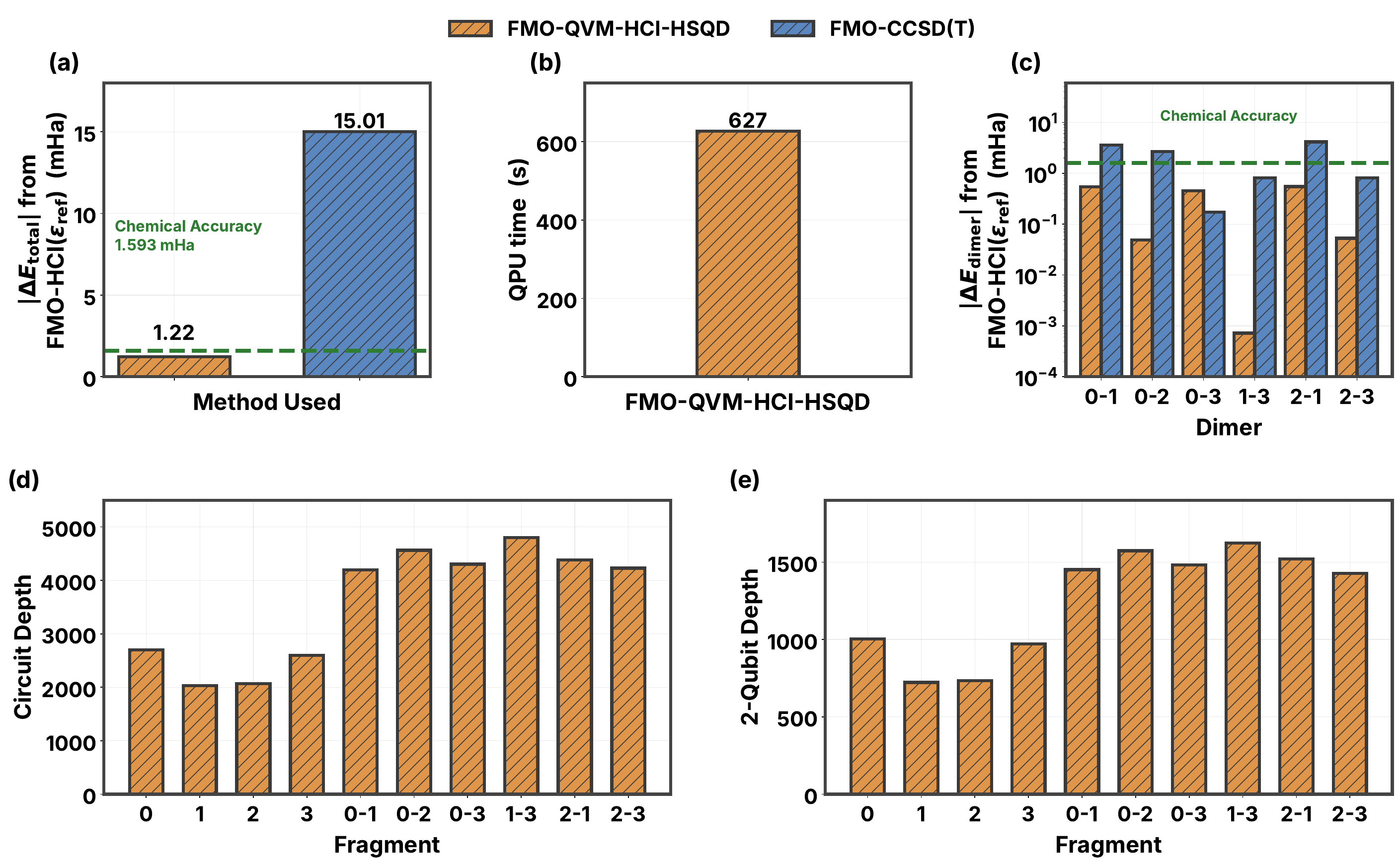}
    \caption{Results for the carmofur--M\textsuperscript{pro} complex calculation in the 6-31G basis. Deviations are computed with respect to FMO-HCI($\varepsilon_{\mathrm{ref}} = 4\times10^{-5}$~Ha). The quantum samples were collected on \texttt{ibm\_boston} with 500{,}000 shots per fragment. (a)~Absolute deviation of the FMO2 total energy for FMO-QVM-HCI-HSQD and FMO-CCSD(T). (b)~QPU time of FMO-QVM-HCI-HSQD. (c)~Absolute deviation of each dimer energy, on a logarithmic scale. (d)~Circuit depth and (e)~two-qubit gate depth of each monomer and dimer circuit. The dashed green line in panels~(a) and~(c) marks chemical accuracy (1.593~mHa).}
    \label{fig:Carmofur_631-g}
\end{figure}

Since the active space is too large for exact diagonalization, an exact FCI reference is not available for this system. We therefore use two approximate references to assess the performance of our algorithm, FMO-CCSD(T) and FMO-HCI. FMO-HCI energy converges systematically towards the FCI limit as the threshold is lowered, because a smaller threshold admits more determinants into the variational space and leaves a smaller part of the correlation energy to the perturbative correction. We therefore run the reference with a lower threshold than the one used in FMO-QVM-HCI-HSQD ($\varepsilon_{\mathrm{var}} = 5\times10^{-5}$~Ha), choosing $\varepsilon_{\mathrm{ref}} = 4\times10^{-5}$~Ha. The reference is seeded only with the Hartree-Fock determinant, so it is independent of the quantum samples. We report the circuit depth, two-qubit depth and QPU time on the hardware side, and the deviations of the total and fragment energies on the accuracy side. The results are shown in Fig.~\ref{fig:Carmofur_631-g}.

PT2 corrections were computed with $\varepsilon_{\mathrm{det}} = 10^{-5}$~Ha, $\varepsilon_{\mathrm{stoch}} = 10^{-8}$~Ha and a stochastic target error of $2\times10^{-4}$~Ha. The classical reference uses identical PT2 settings with a smaller uniform selection threshold of $\varepsilon_{\mathrm{ref}} = 4\times10^{-5}$~Ha for all fragments, and is denoted FMO-HCI($\varepsilon_{\mathrm{ref}}$).

For five of the six dimers, the deviation of FMO-QVM-HCI-HSQD from the reference is smaller than that of FMO-CCSD(T). The exception is dimer 0-3, where FMO-QVM-HCI-HSQD deviates by 0.45~mHa compared with 0.17~mHa for FMO-CCSD(T), which is still well within chemical accuracy. In contrast, FMO-CCSD(T) exceeds chemical accuracy for three of the six dimers, with deviations of up to 4.14~mHa. For the total energy, FMO-QVM-HCI-HSQD deviates from the reference by 1.22~mHa, within chemical accuracy, whereas FMO-CCSD(T) deviates by 15.0~mHa. This comparison should, however, be interpreted with care. Our solver and the reference are both based on HCI, with similar selection thresholds and identical PT2 settings, so the agreement between them does not by itself demonstrate that the quantum samples contribute to the accuracy. To address this, Appendix~\ref{app:seeds} compares HCI calculations seeded with classical and with quantum samples, quantifying the contribution of the quantum samples to the final accuracy. Overall, these results demonstrate a resource-efficient workflow for implementing sample-based quantum algorithms on large molecular systems.
\section{Conclusion}
\label{sec:conclusion}

In this work, we presented FMO-QVM-HCI-HSQD, a resource-efficient framework for sample-based quantum simulation of molecules that combines three components: (i) FMO decomposition, which reduces a large molecule to monomer and dimer problems embedded in the electrostatic field of their environment and solvable independently; (ii) a half-qubit sampling ansatz, which halves the number of qubits and reduces the two-qubit depth, helping the hardware produce relevant bitstrings that are then expanded by HCI while keeping the variational determinant space compact; and (iii) quantum virtual machines, which execute independent fragment circuits concurrently on disjoint, quality-weighted regions of a single QPU.

For the water trimer, with 14 spin orbitals per monomer and 28 per dimer, FMO-QVM-HCI-HSQD lowered the two-qubit depth by 39\% relative to FMO-QVM-SQD and the QPU time by 82\% relative to sequential FMO-SQD. It reached more accurate energies while accessing only 3\% of the determinant subspace, compared with 40\% for FMO-SQD. For the carmofur--M\textsuperscript{pro} complex, with 16 spin orbitals per monomer and 32 per dimer, the method lowered the two-qubit depth by 47\% and the QPU time by 20\% relative to FMO-QVM-SQD, and accessed only 0.007\% of the determinant subspace compared with 40\% for FMO-QVM-SQD, while staying within chemical accuracy. When the active space was extended in the 6-31G basis to 52 spin orbitals per monomer and 104 per dimer, sampled on 26 and 52 qubits respectively, the total energy remained within 1.22~mHa of FMO-HCI, inside chemical accuracy.

Several limitations of the present study should be noted. First, the references for the 6-31G calculations are themselves approximate, so the reported deviations measure agreement with these references rather than with the exact ground state. Second, the half-qubit ansatz samples only one spin sector, and correlation between the two spin sectors is recovered classically by the HCI expansion rather than by the quantum circuit. In addition, the FMO decomposition introduces its own truncation error, which we did not quantify against an unfragmented calculation.
Third, the benefit of QVM depends strongly on hardware capacity. On the 156-qubit device used here, only one 52-qubit dimer circuit could be placed on qubits of sufficient quality at a time, so the six dimers of the 6-31G calculation were executed one after another, and only the monomers gained from concurrent execution. Running many regions concurrently also makes qubit mapping harder, which can increase circuit depth and place circuits on lower-quality qubits, degrading the sample quality. As devices grow in qubit count and quality, we expect QVMs to become considerably more useful, and better qubit-allocation strategies are an exciting direction for future work. Finally, the reported QPU times and energies come from single hardware runs, so the variation between repeated hardware executions was not assessed.


\section*{Data Availability}
The datasets generated during the study are available from corresponding author upon request.

\section*{Acknowledgements}
We acknowledge the use of \textbf{IBM Quantum Credits via the IBM Quantum Startups Program }for this work. The views expressed
are those of the authors and do not reflect the official policy or position of IBM or the IBM Quantum Platform team. \\
We would also like to thank Sarvagya Kaushik (IIT-ISM  Dhanbad) for his contributions in QVM development during his internship time at QClairvoyance Quantum Labs. The authors would like to extend their appreciation to the advisors of Qclairvoyance Quantum Labs for their support.

 \section*{Funding}
No funding was received for this research.

\section*{Competing Interests}
The authors declare no competing interests.

\bibliography{references}

\newpage

\appendix
\section*{Supplementary Information}
\setcounter{table}{0}
\renewcommand{\thetable}{A\arabic{table}}

\setcounter{figure}{0}
\renewcommand{\thefigure}{A\arabic{figure}}

\setcounter{equation}{0}
\renewcommand{\theequation}{A\arabic{equation}}

\section{QVM Fragment-to-Region Allocation}
\label{appendixA}

\begin{figure}[H]
\centering

\includegraphics[width=0.85\linewidth]{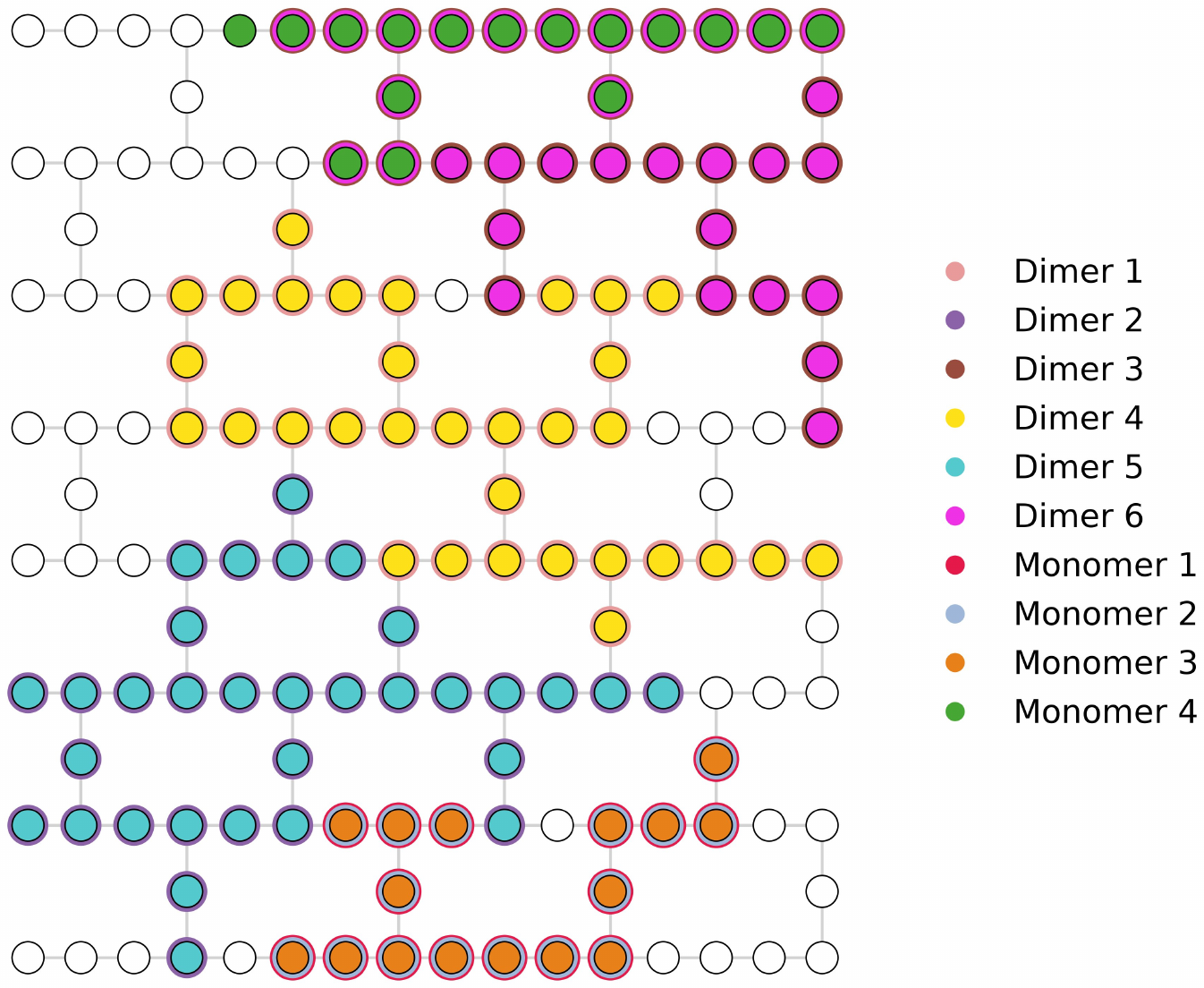}

\caption{QVM-based physical allocation of FMO-SQD fragments on \texttt{ibm\_marrakesh}.}
\label{fig:qvm_fmo_sqd_allocation}

\end{figure}

This appendix gives the fragment-to-region assignments for the carmofur--M\textsuperscript{pro} complex in the STO-3G basis on \texttt{ibm\_marrakesh}. Tables~\ref{tab:qvm_hsqd_allocation} and~\ref{tab:qvm_fmo_sqd_allocation} list the FMO-QVM-HCI-HSQD and FMO-QVM-SQD workloads, and Fig.~\ref{fig:qvm_fmo_sqd_allocation} shows the physical mapping of the FMO-QVM-SQD workload.


\begin{figure}[H]
\centering

\begin{minipage}[t]{0.47\linewidth}
\centering

\captionof{table}{QVM fragment-to-region allocation for the
FMO-HCI-HSQD workload.}
\label{tab:qvm_hsqd_allocation}

\begin{tabular}{c c c}
\hline
\textbf{Fragment} &
\textbf{Qubits} &
\textbf{Region(s)} \\
\hline

Dimer-1 & 16 & $R_{14}$ \\
Dimer-2 & 16 & $R_{5}+R_{10}$ \\
Dimer-3 & 16 & $R_{6}+R_{8}$\\
Dimer-4& 16 & $R_{11}+R_{13}$ \\
Dimer-5& 16 & $R_{14}$ \\
Dimer-6& 16 & $R_{3}+R_{6}$  \\
Monomer-1 & 8  & $R_3$ \\
Monomer-2 & 8  & $R_4$ \\
Monomer-3 & 8  & $R_7$ \\
Monomer-4 & 8  & $R_9$ \\

\hline
\end{tabular}

\end{minipage}
\hfill
\begin{minipage}[t]{0.47\linewidth}
\centering

\captionof{table}{QVM fragment-to-region allocation for the
FMO-SQD workload.}
\label{tab:qvm_fmo_sqd_allocation}

\begin{tabular}{c c c}
\hline
\textbf{Fragment} &
\textbf{Qubits} &
\textbf{Region(s)} \\
\hline
Dimer-1 & 32 & $R_{4}+R_{10}+R_{11}$ \\
Dimer-2 & 32 & $R_{7}+R_{9}+R_{15}$\\
Dimer-3 & 32 & $R_{3}+R_{5}+R_{6}+R_{8}$ \\
Dimer-4 & 32 & $R_{4}+R_{10}+R_{11}$ \\
Dimer-5 & 32 &$R_{7}+R_{9}+R_{15}$ \\
Dimer-6 & 32 & $R_{3}+R_{5}+R_{6}+R_{8}$ \\
Monomer-1 & 16 & $R_{14}$ \\
Monomer-2 & 16 & $R_{14}$ \\
Monomer-3 & 16 & $R_{14}$ \\
Monomer-4 & 16 & $R_{3}+R_{6}$ \\
\hline
\end{tabular}

\end{minipage}

\end{figure}


\section{Quality Comparison of Sampled Bitstrings}
\label{app:sampling}
\begin{figure}[H]
    \centering
    \includegraphics[width=1.0\linewidth]{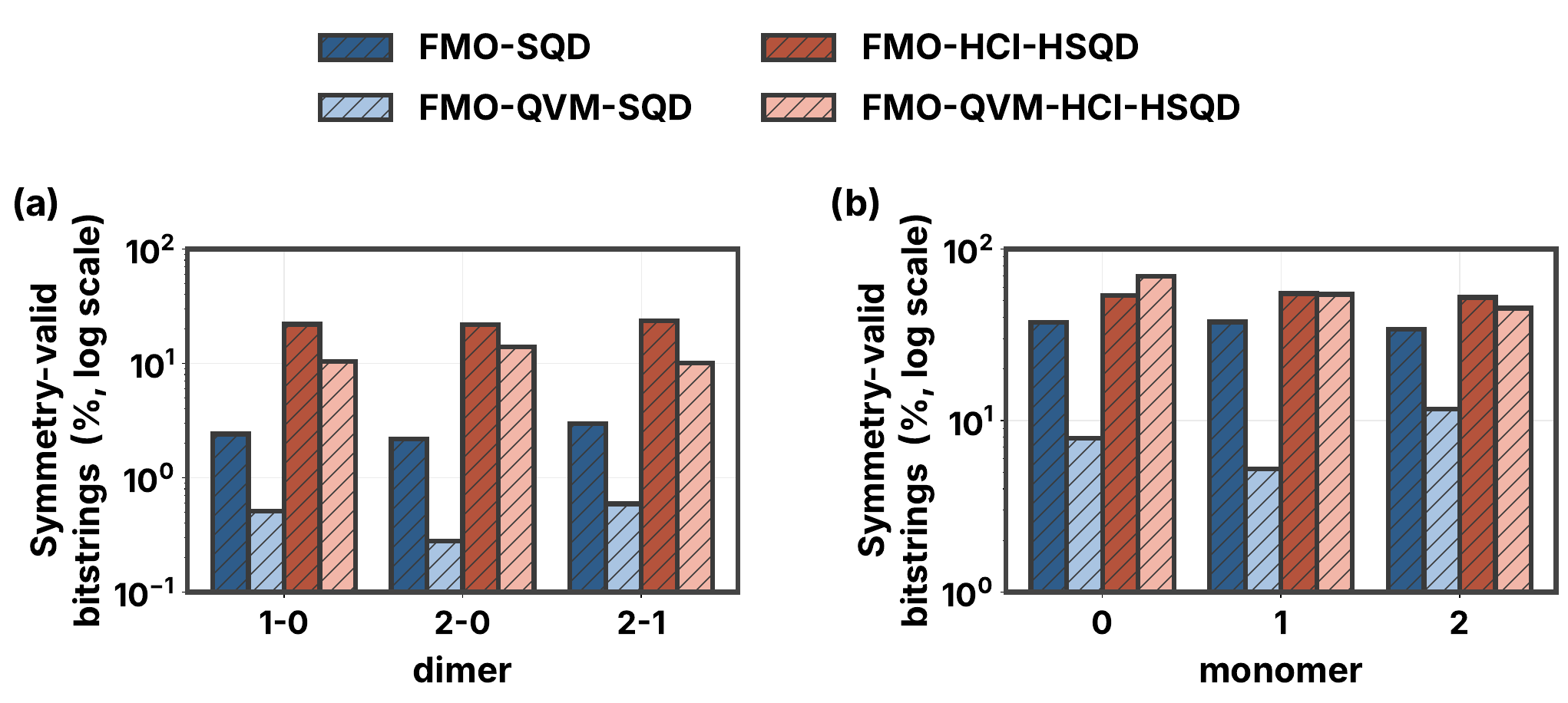}
    \caption{Percentage of symmetry-following bitstrings out of 10{,}000 shots on \texttt{ibm\_marrakesh} for the water trimer, on a logarithmic scale. (a)~Dimers. (b)~Monomers. For SQD, both spin halves must contain the correct number of electrons. For HCI-HSQD, the single sampled spin sector must contain the correct number of electrons.}
    \label{fig:sampling_quality}
\end{figure}

In this section we compare the quality of the bitstrings obtained with the four methods, FMO-SQD, FMO-HCI-HSQD, FMO-QVM-SQD and FMO-QVM-HCI-HSQD, for the water trimer. As a measure of quality, we use the percentage of the 10{,}000 measured bitstrings that satisfy the particle-number symmetry of the fragment, shown in Fig.~\ref{fig:sampling_quality}. For SQD, a bitstring is valid only if both its $\alpha$ and $\beta$ halves contain the correct number of electrons. For HCI-HSQD, each bitstring encodes a single spin sector and is valid if it contains the correct number of electrons in that sector. 
In sequential execution, HCI-HSQD yields 21.9--23.7\% valid bitstrings for the dimers, compared with 2.2--3.0\% for SQD, so roughly nine times more of the measured shots can be used to build the subspace. This gain comes partly from sampling a single spin sector on half the qubits, which reduces the number of bits that must be correct, and partly from the shallower circuits. 

The role of circuit depth becomes clear under QVM, where the routing constraints described in the main text increase the depth of every circuit. For FMO-QVM-SQD, the valid fraction of the dimers drops to 0.28--0.59\%, so the deeper 28-qubit circuits produce almost no usable bitstrings. FMO-QVM-HCI-HSQD retains 10.1--14.0\% valid bitstrings because its circuits remain substantially shallower. The monomers follow the same trend. Under QVM, the SQD monomers drop from 34.0--37.6\% to 5.2--11.6\% valid bitstrings, whereas the HCI-HSQD monomers retain 45.3--69.4\%, close to their sequential values of 52.4--55.1\%. Overall, HCI-HSQD delivers a much larger number of usable bitstrings per shot, both because it samples a smaller space and because its shallower circuits are less affected by hardware noise, and this advantage is most pronounced under QVM execution.

\section{Classical and Quantum Seed Comparison}
\label{app:seeds}
In this section, we assess the contribution of the quantum samples to the final HCI energy by using them as seeds for the HCI expansion. We consider the carmofur--M\textsuperscript{pro} dimer 1-3 in the STO-3G basis, which has 16 spatial orbitals and therefore admits an exact CASCI reference. All calculations use $\varepsilon_{\mathrm{var}} = 10^{-4}$~Ha and report the variational energy only. The HCI expansion is seeded with the Hartree--Fock determinant together with either the 5{,}000 most relevant determinants obtained from the quantum samples or the 5{,}000 most relevant determinants obtained from CISD-ranked strings. As an additional reference, HCI is also started from the Hartree--Fock determinant alone with a smaller threshold of $\varepsilon_{\mathrm{var}} = 5\times10^{-5}$~Ha.

\begin{figure}[H]
    \centering
    \includegraphics[width=1.0\linewidth]{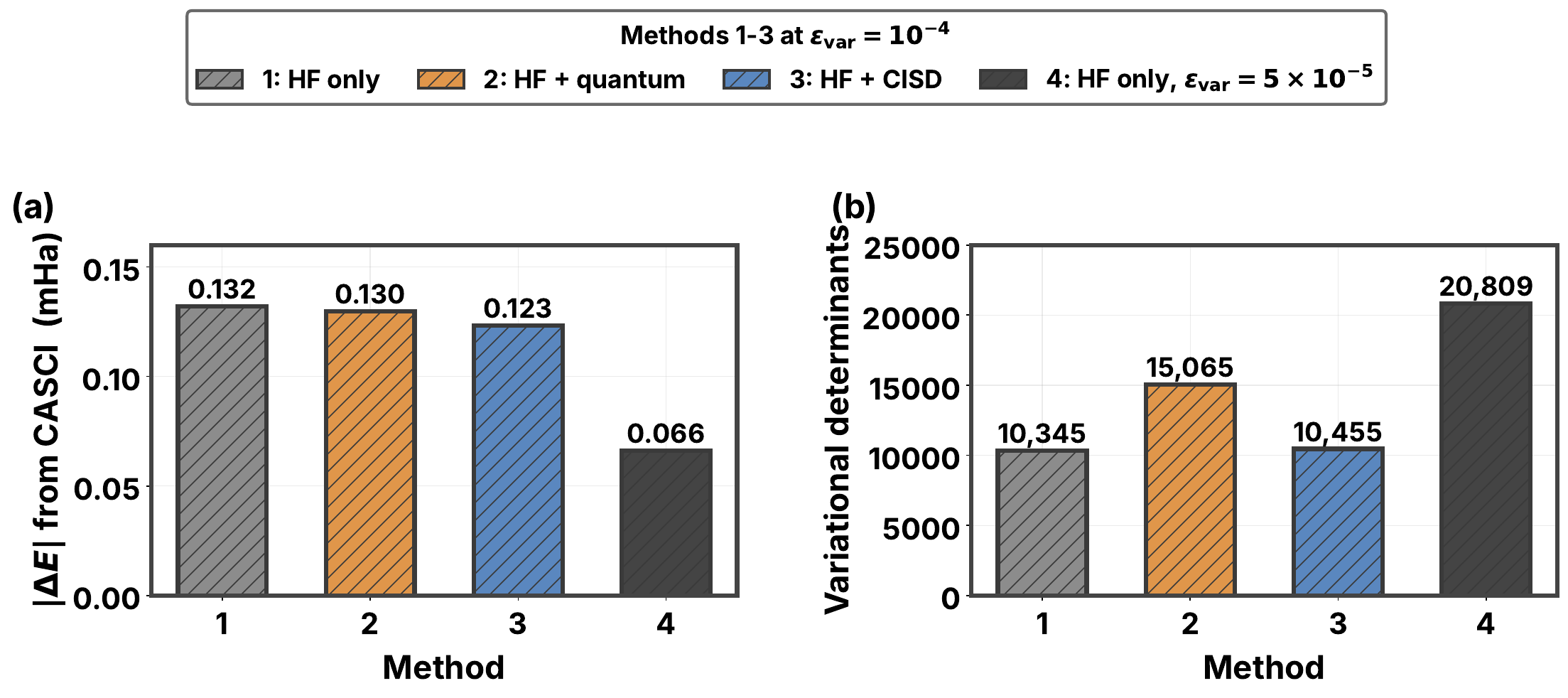}
    \caption{Effect of the initial determinants on HCI for the carmofur--M\textsuperscript{pro} dimer 1-3 (STO-3G, 16 spatial orbitals). Methods 1--3 use $\varepsilon_{\mathrm{var}} = 10^{-4}$~Ha and start from the Hartree--Fock determinant alone (1), or together with the 5{,}000 largest determinants of the $\mathcal{P}\otimes\mathcal{P}$ wave function built from the quantum-sampled spin strings (2) or from an equal number of CISD-ranked spin strings (3). Method 4 starts from the Hartree--Fock determinant alone with $\varepsilon_{\mathrm{var}} = 5\times10^{-5}$~Ha. (a)~Deviation of the variational energy from CASCI. (b)~Number of determinants in the final variational space.}
    \label{fig:seed_comparison}
\end{figure}

The results are shown in Fig.~\ref{fig:seed_comparison}. Starting from the Hartree--Fock determinant alone, HCI reaches a deviation of 0.132~mHa from CASCI with 10{,}345 determinants. Adding the quantum seeds lowers the deviation only to 0.130~mHa, while the final variational space grows by about 46\% to 15{,}065 determinants. The quantum seeds therefore enlarge the space considerably but contribute very little to the energy. In contrast, the CISD seeds lower the deviation to 0.123~mHa with 10{,}455 determinants, a size comparable to that of the Hartree--Fock-only expansion. The CISD-ranked strings are concentrated in low excitations, where most of the correlation energy resides, and therefore guide the expansion towards the relevant part of the determinant space at almost no additional cost. Finally, tightening the threshold to $\varepsilon_{\mathrm{var}} = 5\times10^{-5}$~Ha for the Hartree--Fock-only expansion halves the deviation to 0.066~mHa. Within the HCI framework, the selection threshold therefore controls the accuracy far more strongly than the choice of seeds.

\end{document}